\documentclass[letterpaper,11pt]{article}

\raggedright
\usepackage[svgnames]{xcolor}
\usepackage{framed}
\usepackage{tocloft}
\usepackage{color, soul}
\usepackage{authblk}
\usepackage{graphicx}
\usepackage[T1]{fontenc}
\usepackage{times}
\usepackage{natbib}
\usepackage{ragged2e}
\usepackage{scrextend}
\usepackage[english]{babel}
\usepackage{hyphenat}
\usepackage{textcomp}
\usepackage[utf8]{inputenc}
\usepackage{amsmath} 
\usepackage{amssymb}
\usepackage{bm}
\usepackage{exscale}
\usepackage[mathscr]{eucal}
\usepackage{upgreek}

\newcommand*\samethanks[1][\value{footnote}]{\footnotemark[#1]}

\author[1]{Ningwei Li\thanks{these authors contributed equally to this work}}
\author[1] {Nima Sharifi-Mood\samethanks}
\author[1]{Fuquan Tu}
\author[1]{Daeyeon Lee}
\author[1,2]{ Ravi Radhakrishnan}
\author[3]{Tobias Baumgart\thanks{baumgart@sas.upenn.edu}}
\author[1] {Kathleen J. Stebe\thanks{kstebe@seas.upenn.edu}}

\affil[1]{\small Department of Chemical and Biomolecular Engineering, University of Pennsylvania, 220 South 33rd Street, 311A Towne Building, Philadelphia, PA 19104}
\affil[2]{\small Department of Bioengineering, University of Pennsylvania, 210 S. 33rd St., 240 Skirkanich Hall, Philadelphia, PA, 19104}
\affil[3]{\small Department of Chemistry, University of Pennsylvania, 231 S. 34 Street, Philadelphia, PA 19104}
\begin{document}
\pagenumbering{arabic}
\title{Curvature-driven migration of colloids on lipid bilayers}
\date{}
\maketitle
\justify
\begin{abstract}
Colloids and proteins alike can bind to lipid bilayers and move laterally in these two-dimensional fluids. Inspired by proteins that generate membrane curvature, sense the underlying membrane geometry, and migrate to high curvature sites, we explore the question: Can colloids, adhered to lipid bilayers, also sense and respond to membrane geometry? We report the curvature migration of Janus microparticles adhered to giant unilamellar vesicles elongated to present well defined curvature fields. However, unlike proteins, which migrate to minimize membrane bending energy, colloids migrate by an entirely different mechanism. By determining the energy dissipated along a trajectory, the energy field mediating these interactions is inferred to be linear in the local deviatoric curvature, as reported previously for colloids trapped at curved interfaces between immiscible fluids. In this latter system, however, the energy gradients are far larger, so particles move deterministically, whereas the colloids on vesicles move with significant fluctuations in regions of weak curvature gradient. By addressing the role of Brownian motion, we show that the observed curvature migration of colloids on bilayers is indeed a case of curvature capillary migration, with membrane tension playing the role of interfacial tension. Furthermore, since this motion is mediated by membrane tension and shape, it can be modulated, or even turned ``on'' and ``off'', by simply varying these parameters. While particle-particle interactions on lipid membranes have been considered in many contributions, we report here an exciting and previously unexplored modality to actively direct the migration of colloids to desired locations on lipid bilayers.
\end{abstract}
\pagebreak

\setlength{\parindent}{0.3in}
Curvature generating and sensing proteins are exciting to biologists and materials scientists alike. When such proteins associate with a lipid bilayer membrane, they can be moved about simply by changing the membrane shape \cite{Sorre2011, Zhu2011}. For example, when a tether is pulled from a giant unilamellar vesicle (GUV) whose membrane initially contains sparse concentrations of such proteins, they may concentrate in the tether \cite{Tian2009,Capraro2010}. This remarkable coupling of chemical potential to geometry is not only a key mechanism in membrane-mediated trafficking of proteins in cells, it is also an inspiring and  powerful tool for directing assembly. Could such interactions occur for materials at colloidal length scales?  This would imply that, by simply dictating the geometry of a vesicle, materials like colloids associated with membranes could be moved about from one location to another.  This could enable material organization on lipid bilayers, which is a system of broad significance in biophysics \cite{Mitragotri2015, Petros2010}. However, the motivating system of curvature generating proteins is in some ways quite specific; while there are several groups of proteins that have been identified to be curvature generators and sensors, most membrane-associated proteins do not respond in this way. The proteins are posited to adhere to lipid bilayers through particular modalities that distort the host membrane so that they couple energetically to the underlying membrane geometry \cite{Sorre2011, Zhu2011}.\\ 
~\\
When colloids adhere to membranes, the associated distortion field can, in principle, depend on the underlying membrane geometry. The adhesion of colloids to membranes can occur through various non-specific mechanisms, including electrostatic and hydrophobic interactions, and may also involve specific ligand--receptor interactions \cite{Safinya1999,Beales2012, Fery2003}. At equilibrium, the degree of wrapping of the particle, and the associated ``contact line'' where adhered and un-adhered membrane meet at the particle surface are determined by a balance of the adhesion energies and energies related to membrane bending and tension.  Alternatively, the membrane can become trapped at pinning sites, with an undulated contour at the contact line (Fig.~1a,~1b).  Since contact lines may pin on asperities or patches with high adhesion energy, there can be considerable variation from particle to particle in this case. There is a third possibility; membranes can wrap colloids nearly completely so that they are connected to the outer membrane only via a catenoid-shaped neck \cite{Weikl2014}.  In this state, particle-sourced distortions might be  isolated from the surrounding membrane and effectively decoupled from the membrane geometry around the particle. To explore and develop membrane curvature-mediated particle migration, then, it might be important to limit the degree of particle wrapping.  We use a Janus colloid fabricated via seeded emulsion polymerization which has two faces, one primarily comprising polystyrene (PS), the other  polyacrylic acid (PAA) (Fig.~1c) \cite{Tu2014}. The PAA face of the particle adheres via electrostatic interactions to oppositely charged lipid bilayers, whereas the PS face does not, thereby limiting the degree of wrapping (Fig.~1d). We compare the behavior of these Janus colloids to carboxy-functionalized PS colloids that can be  completely wrapped (Fig.~1e) to explore the importance of this effect.\\
~\\
To determine whether the colloids indeed interact with membrane geometry, we study particles adhered to GUVs that are deformed to impose a well-defined membrane shape. We find that particles move only when the tension is higher than roughly $0.05$~${{mN} \mathord{\left/{\vphantom {{mN} m}} \right.\kern-\nulldelimiterspace} m}$. Furthermore, experiment reveals that particles can dissipate more than $100$~$k_BT$ of energy along a trajectory for the range of tensions over which we study migration, ($0.24-0.69$~${{mN} \mathord{\left/{\vphantom {{mN} m}} \right.\kern-\nulldelimiterspace} m}$). This dependence on tension, and the magnitude of the associated energy changes, imply that these migrations, while inspired by the curvature generating proteins, occur via an entirely  different mechanism. We use experiment and theory to determine this mechanism, and explore features of this new modality for membrane-mediated colloidal assembly. \\

\begin{figure}
\centering
\includegraphics[scale=0.5]{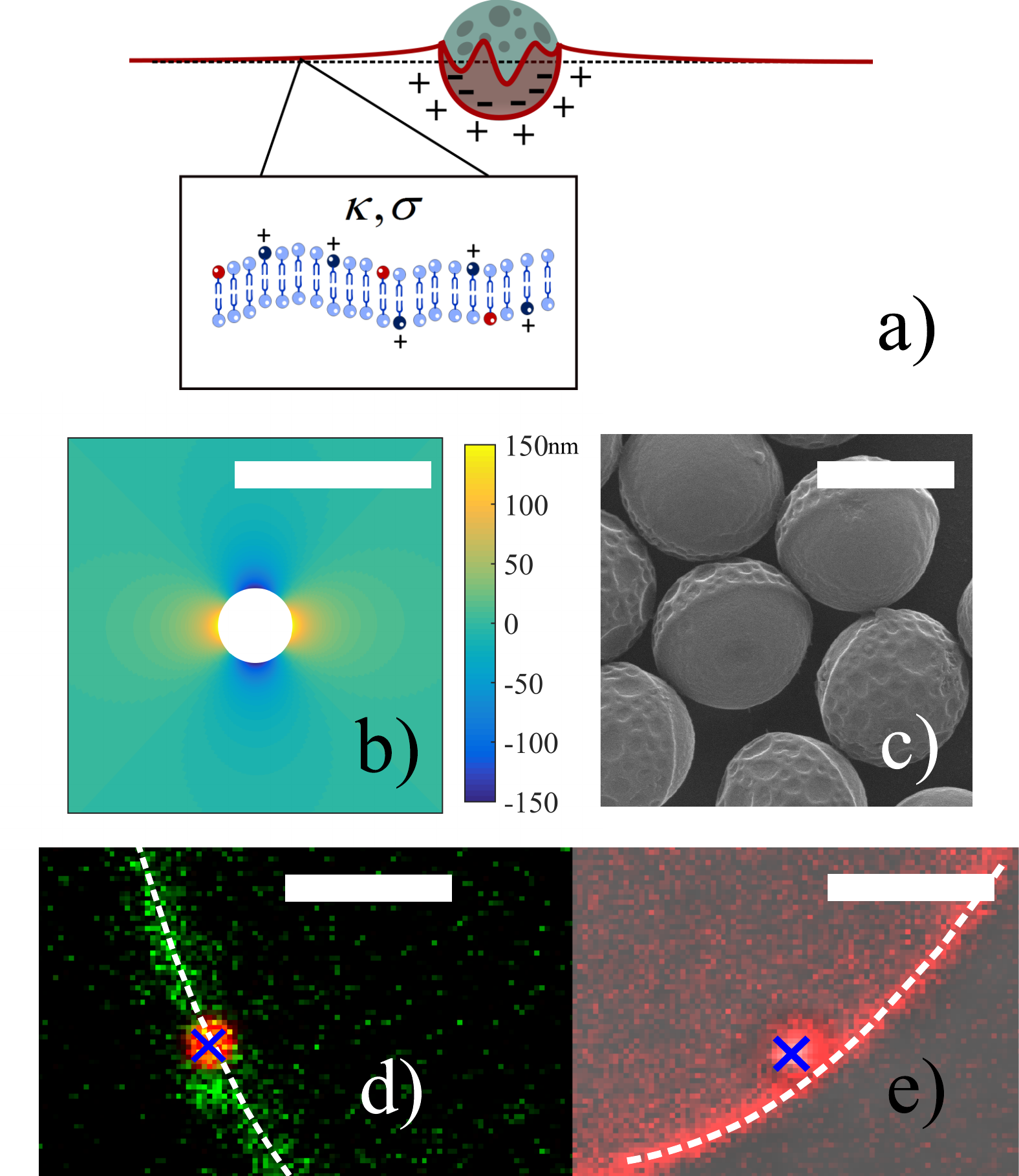}
\begin{addmargin}[.2in]{.2in}
\caption{\footnotesize \textbf{Homogeneous and Janus colloids adhering via electrostatic interactions to a GUV} a) Schematic of a particle partially wrapped by a lipid bilayer with undulated contact line. b) Heat map of the leading order mode in the particle-sourced membrane deformation field, a decaying quadrupolar mode, top view. Scale bar: 2.5 ${\mu}m$. c) Scanning electron microscope image of Janus particles with PAA side (smooth) and PS side with PAA domains (rough). Scale bar: 1 ${\mu}m$. d) \& e) Confocal fluorescence microscopy image of a homogeneous particle (in panel d) and a Janus particle (in panel e) attached to a lipid bilayer. Scale bar: 5 ${\mu}m$. Membrane shape is traced out by white dashed lines, and locations of particles are indicated by blue crosses. In both cases, the membranes curve outward.  These images show that the homogeneous particle is completely wrapped, while the Janus particle is roughly half-wrapped.}

\end{addmargin}

\end{figure}

~\\
\textbf{Results and Discussion}\\
GUVs are formed in a vesicle formation chamber. 
Thereafter, a single vesicle is isolated, protected, and moved from this suspension to a liquid chamber containing microparticles. This allows us to study microparticle migration in a relatively unobstructed fashion, without interference from other vesicles in the field of view. Using a micropipette, a single vesicle is captured and transferred at constant aspiration pressure, and hence tension, to a liquid chamber containing a suspension of microparticles with otherwise identical solution conditions, i.e., 800 $mM$ glucose and 10 $mM$ phosphate-buffered saline (PBS). Once the GUV is introduced to the chamber, particles begin to adhere to the membrane. Concomitantly, the  GUV is deformed using a large, clean, glass bead of radius 10-30 $\mu~m$ glued to a second micropipette. Using micro-manipulators, the bead is  tapped gently against the GUV to promote adhesion to the membrane, and gently retracted to elongate the GUV into a lemon-like shape (Fig.~2a). The entire arrangement is then held fixed. The micropipettes are configured so that the elongated vesicle shape is axisymmetric. A typical contour of such a deformed GUV, captured using confocal fluorescence microscopy, is shown in Fig.~2b. We define a coordinate $(R,Z)$ with $Z=0$ located at the large bead, and $R=0$ on the axis of symmetry. The meridional arclength $s$ along the contour is measured from the bead, as well. The GUV has constant mean curvature $H$. It also has deviatoric curvature $\Delta c$, defined  by the difference between the principal curvatures, which varies with arclength $s$. Gradients in $\Delta c$ are weak near the middle of the GUV, and are steep near the bead and micropipette (Fig.~2c). Vesicle tensions are varied over the range $0.05$~${{mN}/{m}} \leq \sigma \leq 0.69$~${{mN}/{m}}$. \\
~\\
\begin{figure}
\centering
\includegraphics[scale=0.4]{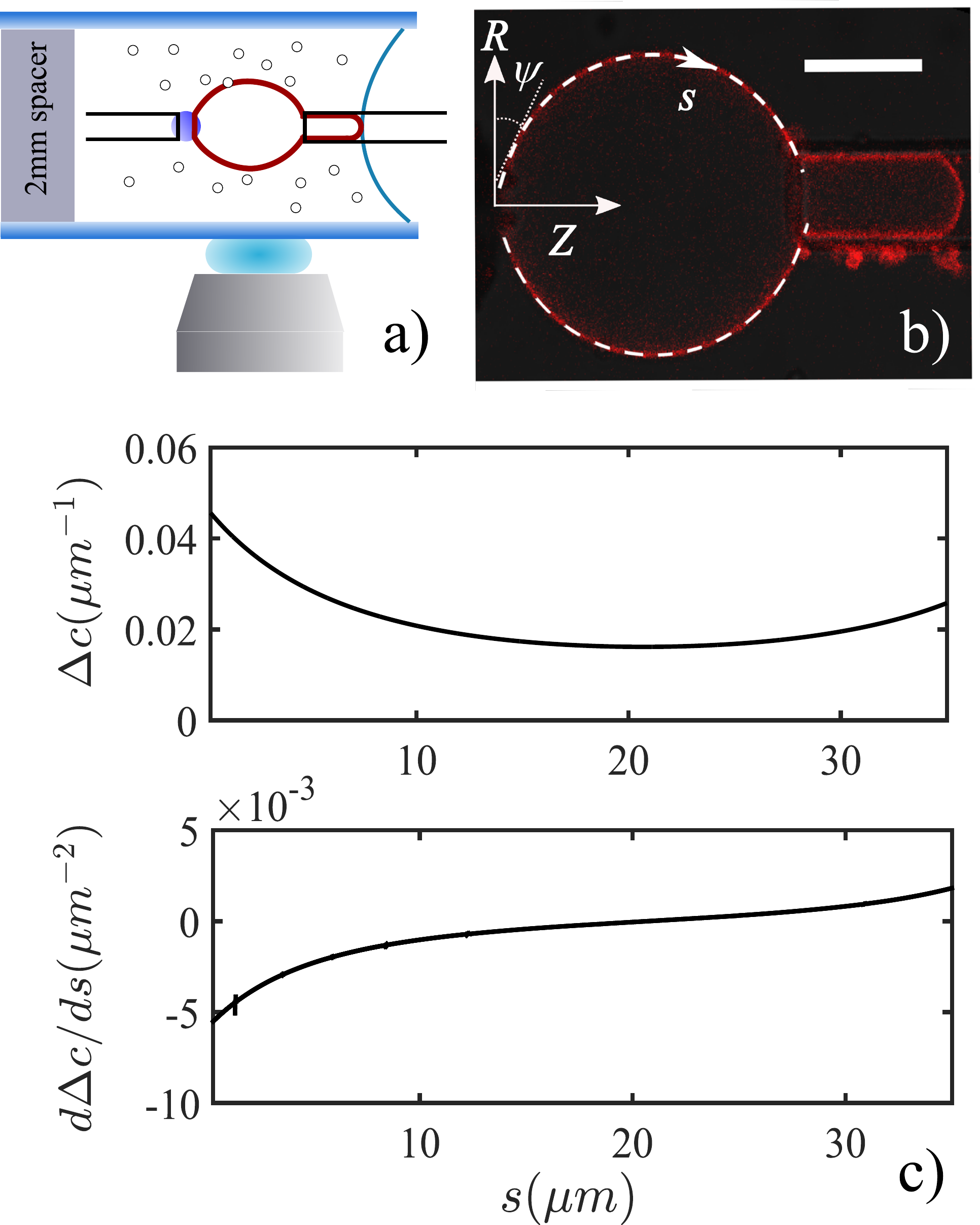}
\begin{addmargin}[.2in]{.2in}
\caption{\footnotesize \textbf{Shape of an elongated GUV} a) Schematic of an aspirated GUV being elongated by two micro-manipulated micropipettes in a liquid chamber with suspended particles. One micropipette holds the aspirated area reservoir. A bead, afixed to the other micropipette, adheres to the GUV and is used to impose elongation. b) Confocal fluorescence microscopy image of an elongated GUV. The white dashed lines show the comparison of the contour obtained by integration of the Young-Laplace equation to the membrane shape. Scale bar: 10 ${\mu}m$. c) Deviatoric curvature (upper panel) and gradient of deviatoric curvature with respect to arc length (lower panel) are plotted against arclength, $s$. These quantities are calculated from the simulated contour.}

\end{addmargin}

\end{figure}

We study GUVs consisting of a mixture of lipids, 40 $\%$ 1,2\hyp{}dioleoyl\hyp{}3\hyp{}trimethylammonium\hyp{}propane(DOTAP) and $59.5~\%$ 1,2\hyp{}dioleoyl\hyp{}sn\hyp{}glycero\hyp{}3\hyp{}phosphocholine(DOPC), and $0.5~\%$ lipid dye (see SI for details) and compare the behavior of two types of particles that adhere electrostatically to this membrane, specifically, isotropic, carboxy-functionalized PS microparticles and the Janus PS-PAA microparticles, both of radius $a=0.5$~$\mu m$. When carboxy-functionalized PS microparticles adhere to the membrane, they become fully wrapped as shown in Fig~1d. On elongated GUVs, over lagtimes of tens of seconds, such particles move with linear mean square displacement (MSD) in arclength $s$ versus lagtime (SI text, Fig.~S1), indicating that this is a nearly diffusion-dominated process, and that the wrapped particle is not strongly coupled to the membrane geometry. \\
The dynamics differ strongly for Janus microparticles. When these particles adhere, confocal microscopy indeed indicates that their degree of wrapping is limited, which we attribute to pinning at the boundary between the two faces. In Fig.~1d, Janus particles, labeled using Nile red, were imaged while adsorbed on the membrane labeled with a green lipid dye, Bodipy-DSPE.  This image indicates that the center of mass of the particle and the lipids are collocated, consistent with the Janus particles being only partially wrapped by the membrane. Once attached to a spherical GUV, the particles have linear MSD, i.e., they move diffusively (see SI text, Fig.~S2). On elongated GUVs, however, the particles move super-diffusively, with motions that can approach deterministic limits, over significant distances to sites of high curvature.  In some cases, particles traverse distances in excess of $30$ particle radii, corresponding to arclengths ranging from $15$ to $20$~ $\mu m$ within $10$ to $15$ seconds. The associated $s$-directed migration velocity $U$ in regions of steep curvature can be so large in regions of steep curvature gradient that particles move with only weak fluctuations, implying that the driving force far exceeds thermal fluctuations (Fig.~3) . This migration is not related to the weak drift velocities in the fluid, as verified by using particles in suspension around the vesicle as tracer particles (see SI text, Fig.~S3). Such trajectories are observed only for tense membranes.   \\
~\\
\begin{figure}
\centering
\includegraphics[scale=0.8]{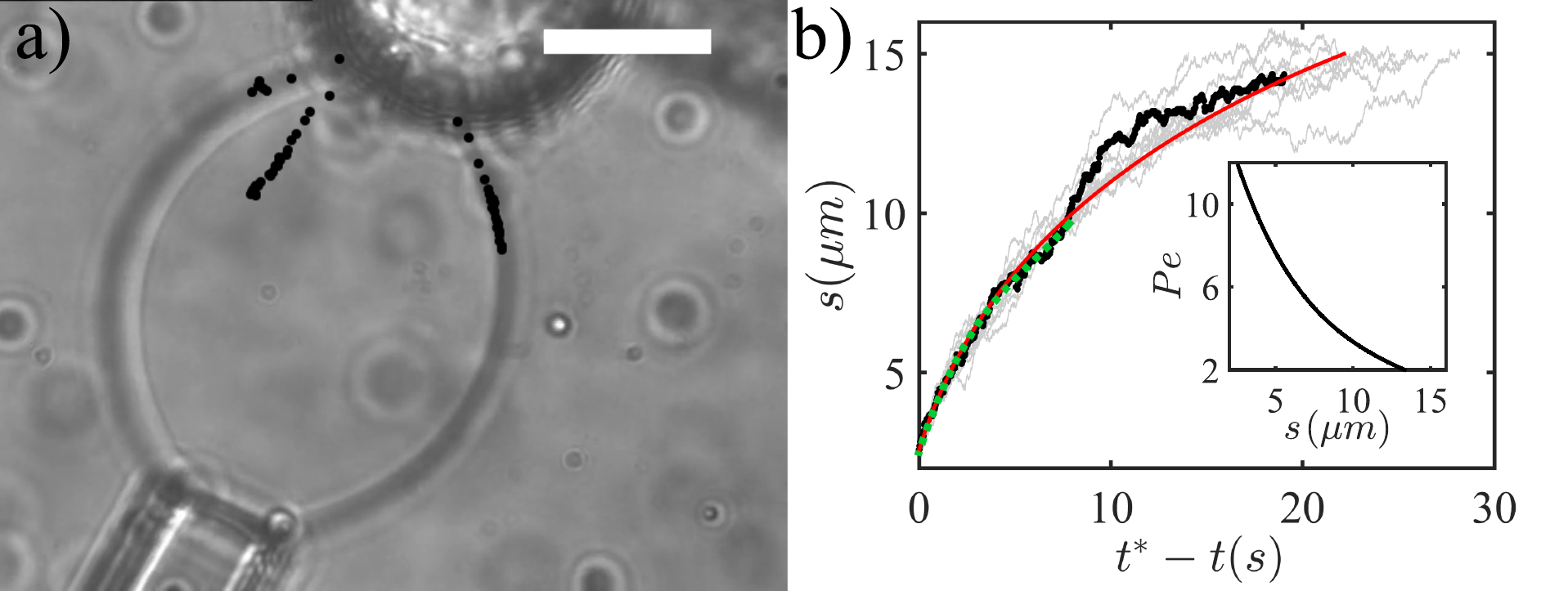}
\begin{addmargin}[.2in]{.2in}
\caption{\footnotesize \textbf{Curvature migration of colloids on an elongated GUV} a) Paths traced by two independently migrating colloids moving along the curvature gradient on an elongated GUV held at fixed tension $\sigma =0.4$~$mN/m$. Particle positions are reported at time intervals of 0.3 $s$.  b) Trajectory of a particle migrating on a GUV held at fixed  tension $\sigma=0.5$~$mN/m$. Black circles: distance of particle from contact with the glass bead in the $s$ direction.  $s(t)$ is plotted against ${t}^{*}-t$, where ${t}^{*}$ is the time that the particle is $5~a$ from  contact with the bead. Green dashed line: cubic fit of the data in the region where $5~a<s<20~a$; red solid line: migration trajectory predicted by imposing a capillary force on a particle with ${{h}_{qp}}=150$~$nm$; solid gray lines: migration trajectories predicted by integration of Langevin equation for $D=0.09$~$\mu {{m}^{2}}/s$ and $T=298$~$K$.}

\end{addmargin}

\end{figure}

We studied vesicles at fixed tensions ranging from $0.05$~$mN/m$ to $0.69$~$mN/m$. For $\sigma \geq 0.24$~$mN/m$, the particles migrate as described above. However, for $ \sigma \sim 0.05$~${{mN} \mathord{\left/
 {\vphantom {{mN} m}} \right.\kern-\nulldelimiterspace} m}$, the particles migrate diffusively (SI text, Fig.~S4). We discuss this transition from migration to diffusive motion at low tension below.  To better understand the forces that drive this displacement, we record the arclength $s$ corresponding to the particle's location at each time step. Particle positions were recorded with a CCD camera at $30$ frames per second. We find  the particles' projected positions on the Z-axis, and use the shape of the elongated GUV to calculate the corresponding particle location in $s$ (see SI text).  Typical particle paths for on the GUV are shown in Fig.~3a, and corresponding trajectories $s(t)$ are plotted in Fig.~3b for paths terminating at the bead. Trajectories are plotted against ${t}^{*}-t$, where ${t}^{*}$ is the time that the particle is $5~a$ from  contact with the bead, to avoid hydrodynamic interactions with the bounding surfaces. Similar results are obtained for particles that migrate toward the micropipette that aspirates the vesicle, owing to the symmetries of the curvature field. \\
~\\
There are two potential sources of energy to drive the observed migration: energies associated with changes in particle adhesion and energy stored in the membrane shape \cite{Saric2012}. For particles with pinned contact lines, only the latter would play a role. If contact lines rearrange, however, the problem becomes far more complex. While we did not observe contact line rearrangement via confocal microscopy, they could, in principle, occur at scales below optical resolution. However, in order for contact line rearrangement to drive the observed migration, a systematic rearrangement of the contact line that is highly coupled to the curvature field would be required, an unlikely occurrence. We do not, therefore, consider this to be the likely mechanism.\\
~\\
The particle trajectories can be analyzed to reveal the dependence of the energy field driving this migration on membrane geometry. The particles move with negligible inertia, i.e., the Reynolds number ${\mathop{\rm Re}\nolimits}  ={{{\rho_p}Ua} \mathord{\left/{\vphantom {{{\rho_p}Ua} \mu }} \right.\kern-\nulldelimiterspace} \mu } \ll 1$ where $\rho_p$ is the particle density, $a$ is the particle radius, and $\mu$ is the solution viscosity. In this limit, energy dissipated along a particle trajectory is balanced by the work performed on the particle by forces driving its motion. Given the noisy trajectory, it is clear that random Brownian forces play a role. The noisiness is most significant in the regions of weak deviatoric curvature gradients. We show below that we can divide regions of the trajectory into Brownian-dominated and near-deterministic regions. In the latter regions, by fitting a polynomial to $s$ vs. $t$ and differentiating with time (Fig.~3b), the velocity $U$ can be determined. Neglecting Brownian contributions in this region, the energy balance on the particle implies:  
\begin{equation}
\Delta E = \frac{{{k_B}T}}{D}\int_{{s_i}}^{{s_f}} {Uds},
\end{equation}
where $D$ is the particle diffusivity on the membrane and the resistance for an isolated particle is given by the Stokes-Einstein relationship. By integrating Eq.~1,  $\Delta E$, the energy dissipated in moving from an initial position $s_i$ to a final position $s_f$, can be inferred for each trajectory. We truncate this integration $5$ radii from boundaries, i.e., either the bead or the micropipette, again, to avoid artifacts associated with hydrodynamic interactions with the bounding surfaces.  We also truncate the integration $20$ particle radii from the boundaries; at distances greater than this value, the motion is diffusive. \\
~\\
We characterized the diffusivity of Janus particles on elongated GUVs. In five cases, particles adhered initially to regions of the membrane with weak curvature gradients, and moved diffusively, allowing $D$ to be determined from the MSD of these particles to be $D~=~0.12 \pm .05$~$\mu {m^2}/s$. For two of these cases, the particles then diffused to regions of steep curvature, where they then migrated.  Thus, we were able to measure the particle's diffusivity before migration, and subsequently, for the same particle, characterize energy dissipated during particle migration.  The particle diffusivity  $D~=~0.07$~$\mu{m^2}/s$ for one trajectory, and $D~=~0.09$~$\mu{m^2}/s$ for another (see SI text, Fig~S2). For these two cases, we report $\Delta E$ in units of $k_BT$ (Fig.~4); $\Delta E > 100{k_B}T$ are found. For all other cases in which we observed particle migration, particles adhered initially to high curvature gradient sites where $U$ was large, precluding measurement of $D$ for those specific trajectories. The product $\Delta E D/k_BT$ is reported for those cases (Fig.~4, inset). Assuming $D~=~0.12$~$\mu {m^2}/s$ for these trajectories, the associated energy dissipated along these trajectories ranges from $50$~$k_BT \le \Delta E \le$ $350$~$k_BT$. For all cases, $\Delta E$ depends linearly on $ (\Delta {c(s_f)}~-~\Delta {c(s_i)})$, the difference in deviatoric curvatures along a particle path from its initial to final position, with  coefficient of linear regression  $R^2 \geq~0.99$. For cases reported in the inset, the slopes vary significantly from trajectory to trajectory. This may be attributable to differences in adhesion state from particle to particle, and associated differences in the magnitudes of particle-sourced distortions and drag coefficients (or diffusivities).\\
~\\
\begin{figure}
\centering
\includegraphics[scale=0.6]{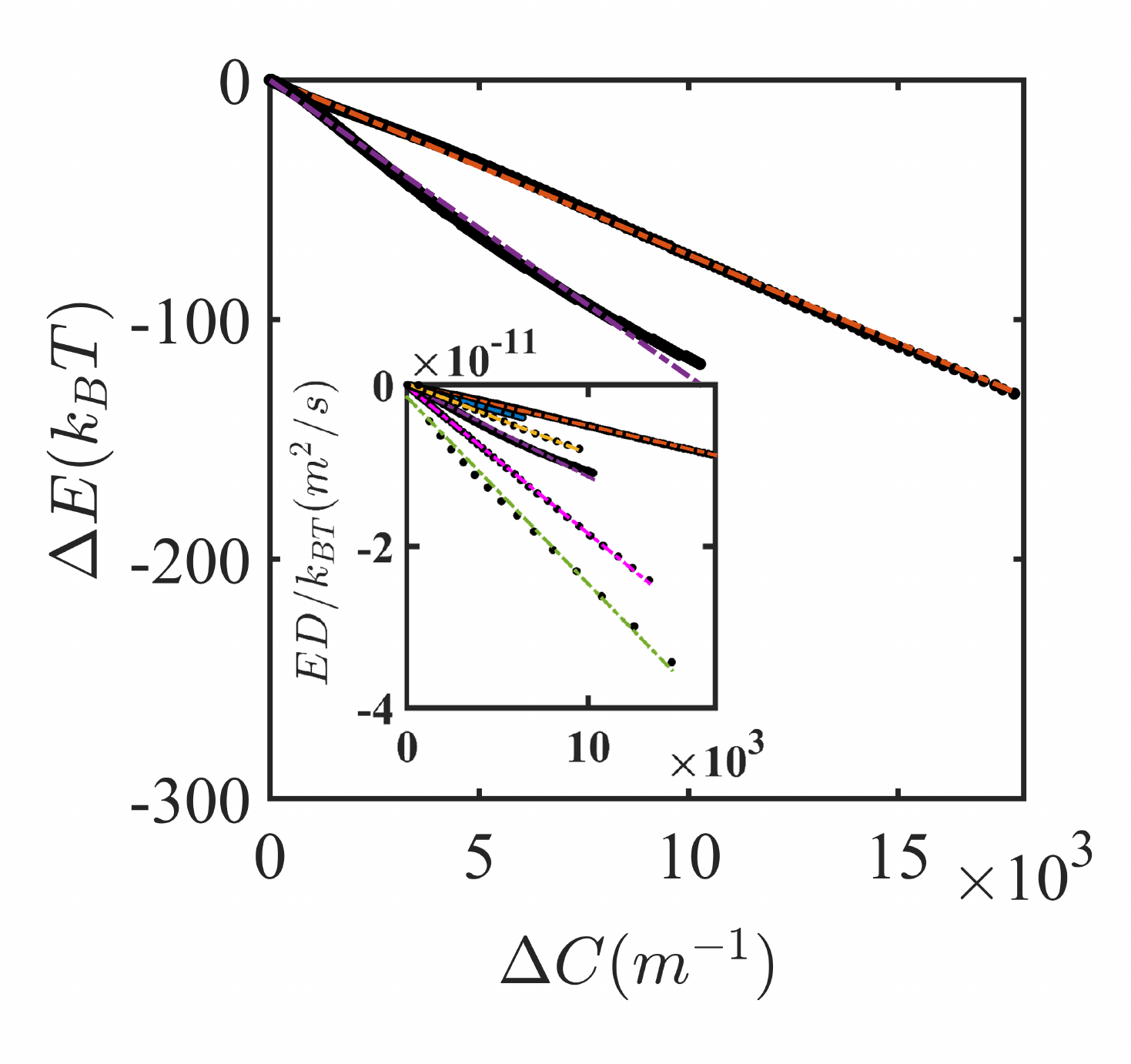}
\begin{addmargin}[.2in]{.2in}
\caption{\footnotesize \textbf{Energy dissipated along the colloids' trajectories}  Figure: the energy dissipated plotted against deviatoric curvature.  These profiles correspond to the two cases for which  particle diffusivity were  measured prior to particle migration. Inset: the energy dissipated, normalized by diffusivity, plotted against deviatoric curvature for all 6 cases for which trajectories were analyzed. $0.24\le \sigma \le 0.69$~$mN/m$.}

\end{addmargin}

\end{figure}

Such a dependence has been reported previously for particles migrating along curvature gradients on interfaces between immiscible fluids. In that case, analysis for the associated change in curvature capillary energy, the product of the interfacial tension and the difference in excess area (created through particle-interface interaction) as it migrates, can be expressed \cite{Cavallaro2011, Liu2015, Mood2016}:  
\begin{equation}
\Delta E =  - \sigma \pi {r_0^2} \frac{h_{qp}\Delta C}{2} , 
\end{equation}
In this expression, for fluid interfaces, $\sigma$ denotes the interfacial tension, and ${{h}_{qp}}$ is the magnitude of the quadrupolar mode of the distortion made by the particle in the interface owing to its undulated contact line located at $r=r_0$, and $\Delta C = \Delta c({s_f})~-~\Delta c({s_i})$ is the change in deviatoric curvature along the particle path. Could the migration of colloids on the tense vesicles be analogous to curvature capillary migration on interfaces between immiscible fluids?  \\ 
~\\
 One can readily develop this analogy, and argue that the particles migrate to reduce excess area in the membrane associated with the distortions made by the particle. There is an important distinction between these systems, however, since membranes can be considered effectively incompressible. How can a particle change the area of an incompressible membrane? The area of the vesicle outside of the pipette can change, since the pipette-aspirated vesicle fraction acts as an area reservoir. By migrating to sites of high deviatoric curvature, the decrease in area of the particle-sourced distortion can return to this reservoir.  To maintain constant tension, the vesicle fraction in the pipette rearranges, lowering the free energy of the system. In this context, we define excess area as the amount of membrane area that is extracted from the area reservoir (represented by the micropipette) at mechanical equilibrium in response to particle binding. In Eqs. 1 and  2, $\Delta E$  represents the change in total internal energy of the system as particle moves from $s_i$ to $s_f$. Since this is a state function, it does not depend on the path of the particle connecting initial and final points on the trajectory. \\
~\\
 To further develop this analogy, the relative importance of bending energy and membrane tension must be clarified in this system. Assuming small slopes and a Monge gauge, the membrane shape around a particle adhered to the membrane, $h(r,\phi)$ is determined by the Euler-Lagrange equation associated with the Helfrich's energy, $\Delta P=\sigma {{\nabla }^{2}}h-\kappa {{\nabla }^{4}}h$ where $\kappa$ is the bending modulus, of typical magnitude of $10$~$k_BT$, and $\Delta P$ is the pressure jump across the membrane.  For tense vesicles, effects associated with $\kappa$  are limited to very small domains adjacent to the particle, of radial extent comparable to the natural length scale $\lambda  = \sqrt {{\kappa  \mathord{\left/{\vphantom {\kappa  \sigma }} \right.\kern-\nulldelimiterspace} \sigma }}$  $\sim 10 - 60$~$nm$.  Outside of these domains, the bending term is negligible, and the membrane shape is determined by the Young-Laplace equation: $\Delta P=\sigma {{\nabla }^{2}}h$.  An asymptotic expansion for $h(r,\phi)$ in small $\epsilon=\lambda/ r_0$ (assuming $r_0 \sim a$) shows that, to leading order, the distortion created by the particle on the curved, tense membrane  is indeed of the same functional form as the disturbance created by a particle on a fluid interface. That is, the undulated ``contact line'' can be decomposed into Fourier modes, each mode exciting a distortion in the surrounding membrane.  The leading order term of this distortion field is a decaying quadrupolar mode, with two contributions on curved interfaces: the particle-sourced distortion, with magnitude $h_{qp}$ and an induced distortion in the interface proportional to the local deviatoric curvature of the host interface.  Furthermore, the leading order excess energy associated with this disturbance can be expressed in the form of Eq.~2, with $\sigma$ denoting membrane tension. Bending contributions are relegated to higher order corrections (see SI text for detail).\\ 
~\\
With this analysis to provide guidance, we can address more carefully the role of Brownian motion, and justify the polynomial fit to the trajectories, and the division of the trajectories into thermally-dominated and near-deterministic regions. A force balance on the particle in the $s$ direction includes the curvature capillary force, given by the negative gradient of the energy expression in  Eq.~2, and a random force owing to thermal fluctuations, expressed by the Langevin equation in the overdamped  state: 
\begin{equation}
\frac{{{k}_{B}}T}{D}\frac{ds}{dt}=-\frac{d\Delta E}{ds}+\sqrt{\frac{2(k_BT)^2}{D}}R(t)
\end{equation}
where $R(t)$ is a random number that has the following characteristics: $<R(t){{>}_{t}}=0$ and  $<R(t)R(t-\tau ){{>}_{t}}=\delta (\tau )$.\\
~\\
Assuming ${r}_{0}=a$, values for $h_{qp}$ of $150$~$nm$ and $100$~$nm$ were inferred from Eq.~2 and Fig.~4 for the two cases in which we were able to fix the diffusivity for the particular trajectories. We integrate Eq.~3 for $h_{qp} = 150$~$nm$, the value that corresponds to the trajectory shown in Fig.~3b. Owing to the presence of the stochastic Brownian term, each integration is different; a set of 15 simulated trajectories, depicted as grey solid lines in Fig.~3b, show distinct zones of behavior. Where curvature gradients are weakest, the simulated trajectories are noisy. In this region, MSD vs. lag time is linear for both experiment and simulation, justifying the identification of this zone as thermally-dominated. Where the curvature gradients are steepest, the predicted and  experimental trajectories converge. For reference in this region, we also show the predicted profile absent thermal fluctuations.  This profile (solid red line) superposes with the converged profiles and the polynomial fit to this region (green dashed line). To quantitatively characterize these regions as thermally dominated or as deterministic in their behavior, we define a local P\'eclet number $Pe = {{U(s)a} \mathord{\left/{\vphantom {{U(s)a} D}} \right.\kern-\nulldelimiterspace} D}$ along a trajectory, defined in terms of the migration velocity of the particle $U(s)$;  $Pe \gg 1$ indicates that the curvature force plays a strong role in the particle motion. In experiment and simulation, $4 \leq Pe \leq 30$ over the range of arclengths where the realizations converge in Fig.~3b. This corresponds to the range of arclengths over which $\Delta E$ was estimated in Fig.~4. \\
~\\
Particles failed to migrate to sites of high curvature for $\sigma=0.05$~${{mN} \mathord{\left/
 {\vphantom {{mN} m}} \right.\kern-\nulldelimiterspace} m}$; for this small tension, the capillary force was always weaker than the Brownian force over the entire trajectory. Thus, the failure to migrate at low tension simply represents a weakening of the capillary force to negligible values (see SI text, Fig.~S5). \\
~\\
Nanoparticles \cite{Yu2009} and proteins \cite{Zhu2011,McMahon2005,Simunovic2015} interact with lipid bilayers, and can, through collective interactions, drive pronounced changes in vesicle morphologies including pearling, and tubule formation.  Since a particle-sourced disturbance has magnitude comparable to the particle radius and decays over similar distances, nanoscopic particles with $a \le \lambda$ are not anticipated to respond to the curvature field via the capillary curvature mechanism explored here.  Nor are curvature generating and sensing proteins, which are also thought to generate disturbances with magnitudes and decay lengths comparable to $\lambda$. The exploration of curvature migration on tense membranes for particles or protein aggregates with sizes between tens of nanometers and microns, the implications for collective interactions and associated vesicle morphologies remains an important and open question for future research.\\
~\\
This mode of interaction opens exciting possibilities for active control of particle assembly.  Since the force is inherently coupled to vesicle curvature, the motion can be  modulated by changes in the membrane shape. We demonstrate this ability by following trajectories of a particle on a vesicle whose shape is dynamically tuned.  When the vesicle is elongated, the particle moves. When the elongation is removed, the particle stops (Fig.~S7).  Migration velocities are, as expected, weak in regions of weak curvature gradient and rapid where gradients are steepest. Are other effects associated with  capillary migration observed? On interfaces between isotropic  fluids, particles interact and assemble. We report such a pair interaction  on a tense membrane in Fig.~5a. However, since membrane tension is much less than typical interfacial tensions (\textit{e.g.} $72$~${{mN} \mathord{\left/{\vphantom {{mN} m}} \right.\kern-\nulldelimiterspace} m}$ for the air-water interface), and the colloids here have relatively small diameters compared to those studied at isotropic interface, these interactions occur only when particles are within roughly 10 particle radii and cannot be interpreted simply in terms of the long range quadrupolar mode \cite{Stamou2000, Loudet2005, Lewandowski2008}. On curved fluid interfaces with finite deviatoric curvature, anisotropic particles rotate owing to curvature capillary torque to align along the principal axes. To explore this phenomenon, we exploit two Janus particles paired to form a dumbbell structure.  Once this dumbbell adheres, we elongate the GUV (Fig.~5b).  The dumbbell rotates at roughly fixed position, and then translates, maintaining its alignment along the meridional direction, with rate of translation increasing with the local curvature gradient. This behavior is indeed quite similar to that observed on isotropic fluid interfaces \cite{Cavallaro2011}.\\
~\\

\begin{figure}
\centering
\includegraphics[scale=0.7]{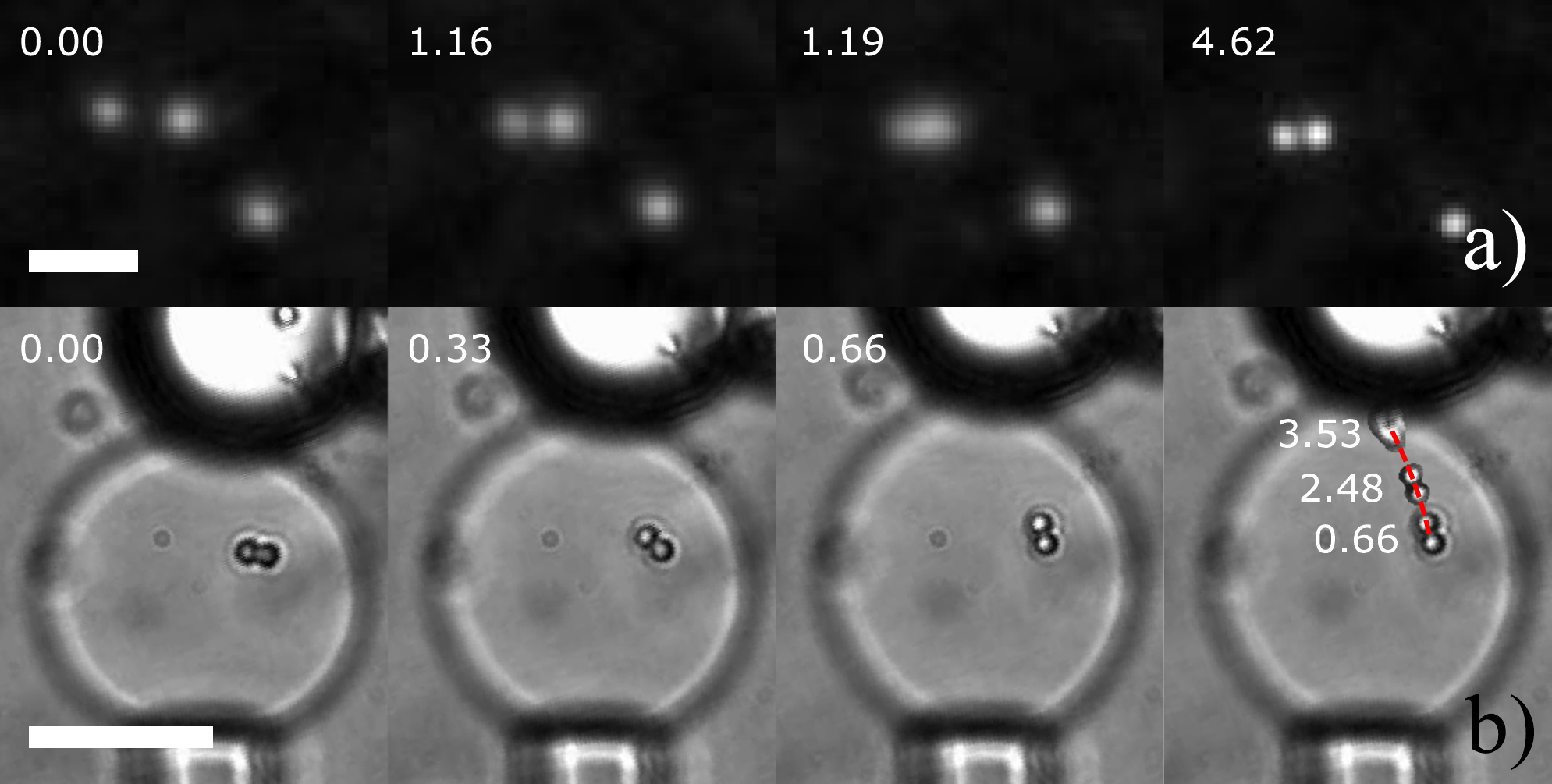}
\begin{addmargin}[.2in]{.2in}
\caption{\footnotesize \textbf{Pair interaction (Top) and Evidence of curvature alignment of anisotropic colloid (Bottom)}. a) Snapshots of an interacting pair of Janus particles on the vesicle. Elapsed time is labeled in seconds in the images. The focus was re-adjusted in the fourth panel. Scale bar: $5$~${\mu}m$. b) First 3 panels: snapshots of a pair of Janus beads, forming a dumbbell shaped dimer, rotating at roughly fixed position to align along the meridional direction. Elapsed time is labeled in seconds in the image. Fourth panel: time-stamped images reporting the location of the dumbbell after this rotation.  Red curve: path traced by dumbbell. The dumbbell migrates while maintaining its orientation. Time is labeled in seconds adjacent to the image of the dumbbell. Scale bar: $15$~${\mu}m$}

\end{addmargin}

\end{figure}

Finally, colloids at interfaces have been assembled in structures resembling a square lattice, consistent with  weak, quadrupolar modes on surfaces with weak curvature gradients \cite{Ershov2013} , and more complex structures, guided by curvature field gradients \cite{Cavallaro2011, Liu2015}. This has exciting implications for many particle interactions on vesicles, which will be the focus of our future studies.\\
~\\
To conclude, we have shown that micron scale colloids, like curvature generating and sensing proteins, adhere to lipid bilayers and migrate to sites of high curvature.  We hypothesize that both particles and proteins ``generate'' a distortion via their adhesion state, and ``sense'' membrane geometry.  We find, however, that particles move by a mechanisms quite different than that of proteins, with nearly deterministic trajectories on tense membranes. The dependence on membrane tension suggests that this interaction is related to the elimination of excess area of the membrane, not to minimization of curvature energy, as in the protein migration case.  We confirm that this migration is a form of curvature capillary migration, with membrane tension playing the role of interfacial tension, and demonstrate the versatility of this previously unexploited modality for colloidal interaction on lipid membranes. \\
~\\
\textbf{Materials and Methods}\\
\textbf{GUV formation} GUVs were formed by a standard electroformation method \cite{Mathivet1996}. Lipids of desired compositions (40 \scalebox{0.75}{$\%$} DOTAP, 59.5 \scalebox{0.75}{$\%$} DOPC and 0.5 \scalebox{0.75}{$\%$} lipid dye) were mixed in chloroform at a total concentration of 1 \scalebox{0.75}{$mM$}, and 40 \scalebox{0.75}{$\mu L$} of the mixture was deposited on each indium tin oxide (ITO) glass slide.  After solvent evaporation in a vacuum chamber for two hours, chambers formed by sandwiching a silicon spacer between two ITO slides were hydrated by 430 \scalebox{0.75}{$\mu L$} of 700 \scalebox{0.75}{$mM$} sucrose solutions and subjected to an alternating current of 0.5 \scalebox{0.75}{$A$} and 10 \scalebox{0.75}{$Hz$} for two hours. Prior to  transfer, GUVs are diluted 20 fold in 850 \scalebox{0.75}{$mM$} glucose with 10 \scalebox{0.75}{$mM$} PBS buffer.\\
~\\
\textbf{GUV shape and membrane tension analysis} The pressure drop across the aspiration pipette was controlled by the height of a water reservoir connecting to the pipette. The height of the water reservoir was monitored by a pressure transducer measuring the hydrostatic pressure of the water reservoir (DP-1520, Validyne, Northridge CA). The contour of the GUV was tracked from both confocal and bright field images using ImageJ and Matlab as shown in Fig.~2b. Under high membrane tension, the shape of a elongated GUV obeys the Young-Laplace equation. An objective function calculating the difference between the experimental contour and the solution of the Young-Laplace equation was minimized in order to find the curvatures of the interface. Once the mean curvature is determined, the tension of the membrane can be also be determined from the Young Laplace equation: 
\begin{equation}
\sigma =\frac{\Delta P}{2}\frac{{{R}_{P}}}{1-{{R}_{P}}H}\
\end{equation}
where \scalebox{0.75}{$\sigma$} is the membrane tension, \scalebox{0.75}{$\Delta P$} is the pressure drop across the micropipette, \scalebox{0.75}{${{R}_{P}}$} is the inner radius of the micropipette, and \scalebox{0.75}{$H$} is the mean curvature of the elongated GUV. The principal curvatures, i.e., the curvatures in the parallel and meridional directions, respectively, of the elongated GUV were also calculated from the fitting results:
\begin{equation}
{{c}_{p}}=\frac{\sin \psi }{R}
\end{equation}
\begin{equation}
{{c}_{m}}=\frac{d\psi }{ds}
\end{equation}
~\\
More details of materials and methods can be found in the SI text, including all chemicals used; GUV transfer and elongation; image acquisition and analysis; sample preparation for SEM and AFM; determination of particle diffusivity; zeta potetial measurement; and derivation of the energy expression of a particle on tense lipid membranes.

\newpage
\setcounter{figure}{0}
\setcounter{equation}{0}
\setcounter{page}{1}
\renewcommand{\thefigure}{S\arabic{figure}}
\renewcommand{\thetable}{S\arabic{table}}
\renewcommand{\theequation}{S\arabic{equation}}
\pagenumbering{roman}

\title{\large Supporting Information}
\maketitle

\textbf{Materials}\\
Lipids including 1,2\hyp{}dioleoyl\hyp{}sn\hyp{}glycero\hyp{}3\hyp{}phosphocholine (DOPC), 1,2\hyp{}dioleoyl\hyp{}3\hyp{}tri\-methyl\-ammonium\hyp{}pro\-pane (DOTAP) were purchased from Avanti Polar Lipids (Alabaster, AL). Lipid dyes N-(4,4-difluoro-5,7-dimethyl-4-bora-3a,4a-diaza-s-indacene-3-propionyl)-1,2-dihexadecanoyl-sn-glycero-3-phosphoethanolamine (BODIPY FL DHPE), triethylammonium salt, and 1,2-dihexadecanoyl-sn-glycero-3-phosphoethanolamine (Texas Red), triethylammonium salt were purchased from Invitrogen (Carlsbad, CA). Glass beads 10-30~$\mu$m and Polybead carboxylate microspheres 1~$\mu$m were obtained from Polysciences, Inc. (Warrington, PA). D-sucrose, dextrose (D-glucose) anhydrous, phosphate buffered saline concentrate powder, and chloroform were purchased from Fisher Scientific (Hampton, NH). Poly-l-lysine 1 $mM$ solution and bovine serum albumin were purchased from Sigma-Aldrich. 5 Minute Epoxy was purchased from Devcon (Danvers, MA).\\
~\\
\textbf{GUV transfer and elongation}\\
GUV transfer, elongation and imaging were performed in two observation chambers formed between two microscope cover slips. The cover slips were separated by a 2 $mm$ thick spacer, and pre-treated by $5~mg/ml$ BSA solution in 1XPBS buffer, followed by incubating with $1~mM$ poly-l-lysine solution for 20 minutes. $350~\mu L$ of background solution ($800~mM$ glucose and $10~mM$ PBS) was deposited in each chamber. $15~\mu L$ of vesicle stock solution was added to one chamber, and $2~\mu L$ of particle stock solution was added to the other. The two chambers were separated by a silicon spacer. \\
~\\
Once formed, a vesicle is selected, aspirated using a glass micropipette, and transferred to the particle chamber. The glass micropipettes used for this purpose were fabricated by pulling $1~mm$ glass capillaries with a brown-flaming micropipette puller (P-77, Sutter Instrument Co., San Francisco CA), and subsequently cut open at desired opening radii by a microforge controller (DMF1000, World Precision Instruments, Sarasota FL).  Thereafter, micropipettes were pre-treated by BSA and poly-l-lysine using methods described above, and used to aspirate the selected vesicle. The aspirated vesicle and micropipette were then covered by a protective "sleeve", a $1.5~mm$ capillary, and transferred to the particle chamber at a constant aspirating pressure along with a small amount of liquid from the vesicle chamber. The sleeve was then removed, leaving the aspirated vesicle in a chamber filled with microparticles.\\
~\\
To deform the vesicle, a $10-30~{\mu}m$ glass bead was glued to the walls of a second glass micropipette using epoxy glue. The glass bead was brought in contact with the vesicle. The membrane adhered to the glass bead strongly. The bead was subsequently moved while maintaining this adhesion so that the vesicle was deformed. The positions of both the aspiration pipette and pipette adhered to the glass beads were controlled by a 3-axes motor system (mini 25, Luigs $\&$ Neumann, Germany).  The position of the bead as it adhered to the vesicle and as it was moved were controlled to impose  an axisymmetric shape on the elongated GUV.  \\
~\\
\textbf{Image acquisition and data analysis}\\
Fluorescence images were obtained by a confocal fluorescent microscope (IX81, Olympus, Japan) with a 60x 1.1 NA water immersion objective (LUMFL, Olympus, Japan). The bright field images were taken by a CCD camera mounted on the same microscope (XC-ST30, Sony, Japan). Particle motion was  captured by the CCD camera at a frame rate of $0.033~second/frame$. Particle position was tracked by ImageJ and further processed by Matlab.\\
~\\
\textbf{Homogeneous particles move diffusively }\\
We also investigated the dynamics of homogenously functionalized PS particles adhered to tense lipid membranes presenting curvature gradients. Particles are 1 ${\mu}m$  carboxylic functionalized PS particles purchased from Invitrogen. The position of a particle attached to an elongated GUV under tension of $0.3~mN/m$ was tracked, and the distance of the particle from the glass bead was determined at each time point (Fig.~S1a); the particle moved with no preferred direction, either toward or away from the bead. The MSD of this trajectory is also linear (Fig.~S1b), indicating that the homogeneous particle moves in a Brownian manner. We calculate the deviatoric curvature difference over the range of $s$ traversed on the elongated GUV, $\Delta c(s_f=4$~$\mu m)-\Delta c(s_i=8$~$\mu m)=0.0132$~$\mu {m}^{-1}$. According to Eq.~2 in the main text, such a curvature difference corresponds to an energy difference of $50~k_B$T for a particle that makes a distortion with the magnitude of $h_{qp}$ of $150~nm$. Therefore, under the same conditions, a typical Janus particle would have migrated. \\
~\\

\begin{figure}
\centering
\includegraphics[scale=0.7]{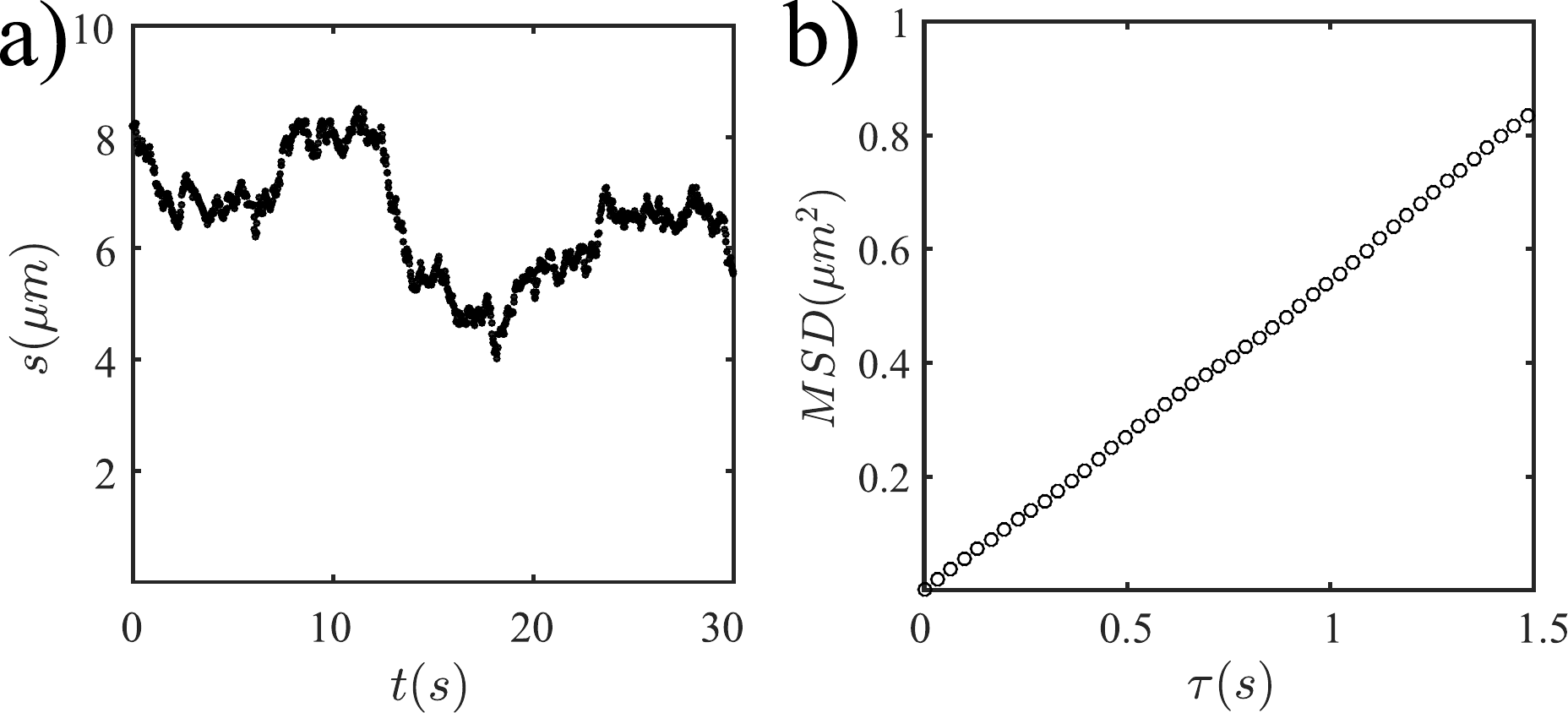}
\begin{addmargin}[.2in]{.2in}
\caption{\footnotesize \textbf{Trajectory of a homogeneous PS colloid adhered to an elongated GUV }a) Trajectory of particle. b) MSD calculated from the trajectory shown in a. } 

\end{addmargin}

\end{figure}

We attribute this failure to respond to the curvature gradient to the wrapping state of the homogeneous particle. Indeed, confocal microscopy reveals that the particle is strongly wrapped, as shown in Fig.~1e in the main text. This suggests that strongly wrapped particles do not deform the surrounding membrane in the same manner as do partially wrapped particles, and the wrapped particles are effectively isolated from the capillary forces associated with the membrane curvature.\\
~\\
\textbf{Diffusivity measurement on elongated GUVs}\\
When a Janus particle is in contact with membranes presenting sufficiently weak curvature gradients, the capillary force acting on the particle is also weak, and thermal fluctuations can be the dominating driving force for particle motion. We were able to measure the diffusivities of isolated Janus particles attached to elongated GUVs for $5$ cases. Janus particles attached to elongated GUVs were imaged by a CCD camera for $300~$frames at a rate of 30 frames per second and tracked using ImageJ. Their projected positions on the $Z$-axis of the elongated GUVs were found from the tracked positions. Displacements on the $s$-coordinate were calculated by the relation:
\begin{equation}
\sin \psi =\frac{dZ}{ds}
\end{equation}
The coordinate system is shown in Fig.~2b in the main text.
The MSDs were calculated from the displacements in $s$-direction and plotted against lag-time (Fig.~S2a). From the slopes, the diffusivities were obtained:
\begin{equation}
D=\frac{MSD}{2\tau}
\end{equation}
Systematic error from noise in the image owing to  imperfection in illumination and image acquisition can contribute to the MSD in the following form \cite{Doyle2005}:
\begin{equation}
\left\langle \Delta {{x}^{2}}_{measured}\left( \tau  \right) \right\rangle =\left\langle \Delta {{x}^{2}}_{real}\left( \tau  \right) \right\rangle +{{\delta }^{2}}
\end{equation}
where $\delta$ is the systematic error. We measured this error by tracking particles fixed on a glass slide in similar illumination environments. The variance of the apparent position of the fixed particles is 0.003~$\mu{{m}^{2}}$ in both $x$ and $y$ directions. For the smallest diffusivity we found, $i.e.$ $D=0.08~\mu m^2/s$, this error is $15~\%$ or less of the displacement of the particle for lag times greater than $0.132~$seconds. Therefore, in order to determine the diffusivity, we consider only data acquired at larger lag times. \\
~\\
The resulting diffusivity is $0.115~\mu {{m}^{2}}/s$ on average, and has a standard deviation of $0.049~\mu {{m}^{2}}/s$. The elongated GUVs were held under various tensions, but no correlation was observed between diffusivity and tension (Fig.~S2b). It has been reported that quantom dots attached to GUVs via biotin-streptavidin linkage can have a wide range of diffusivity\cite{Domanov2011}. While our particles differ in length scale, we also measured a wide range of diffusivities, which might be attributed to differing degrees of wrapping, and associated differing degrees of immersion in the surrounding fluids \cite{Stone2015}, fluctuations in GUV compositions, and impurities such as small lipid vesicles connecting to the GUV. \\
~\\
\begin{figure}
\centering
\includegraphics[scale=0.7]{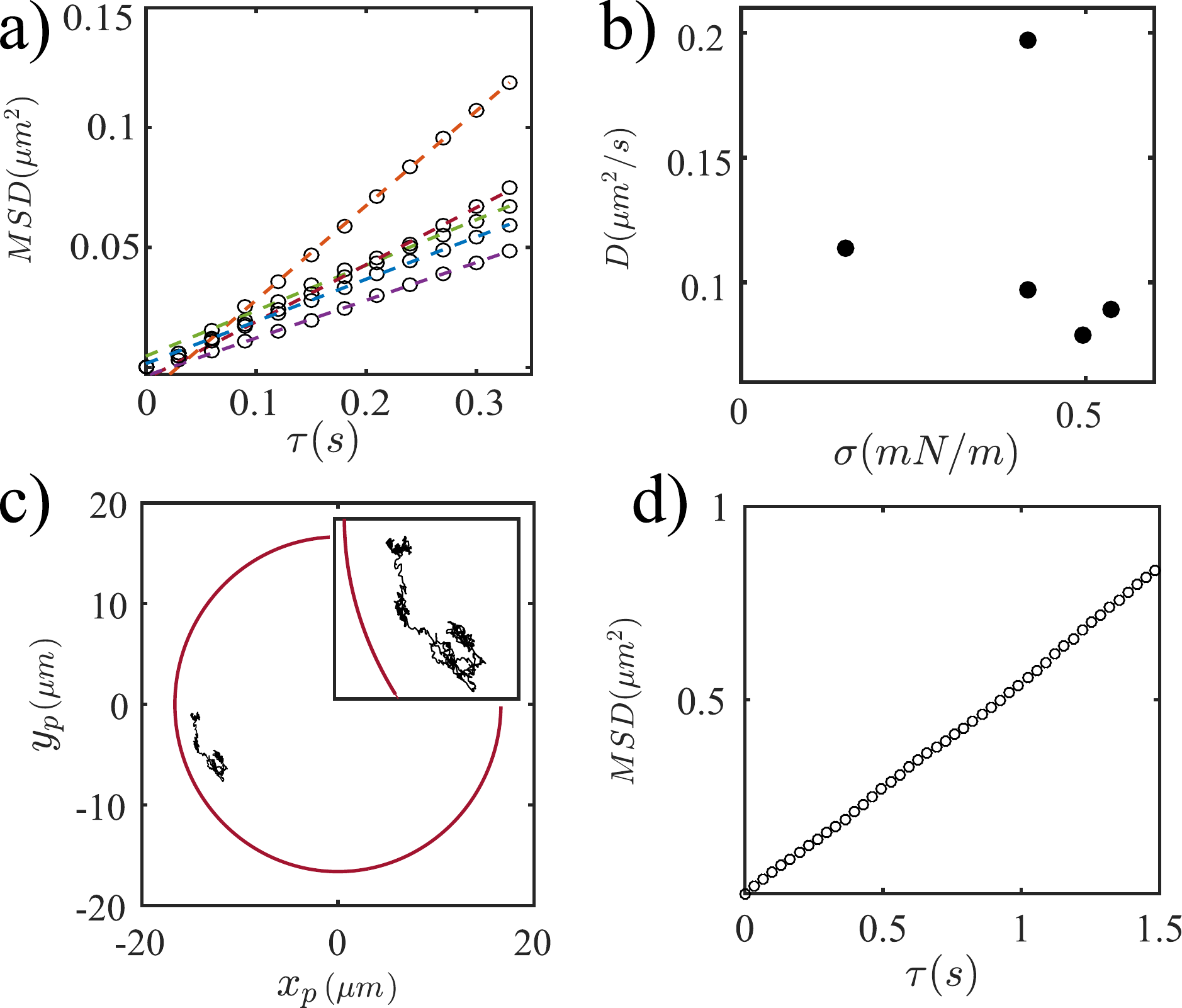}
\begin{addmargin}[.2in]{.2in}
\caption{\footnotesize \textbf{Determination of diffusivities of Janus particles adhered to elongated GUVs, and diffusive trajectory of a Janus particle on a spherical GUV} a) MSDs vs. lag time of Janus particles adhered to the elongated GUVs at regions with a weak curvature gradient. b) Diffusivity of Janus particles on elongated GUVs plotted against membrane tension $\sigma$. c) Typical projected trajectory of a Janus particle on a spherical aspirated GUV indicating by the red circle. The position is shifted so that the coordinate origin is at center of the GUV. Inset: zoomed-in view of the trajectory. d) MSD of the trajectory shown in c. } 

\end{addmargin}

\end{figure}

\textbf{MSD measurement for Janus particles on spherical GUVs}\\
Micropipette aspirated GUVs were transferred to chambers containing Janus particle suspensions. Once particles adhered, the GUV-bound particles were imaged by a CCD camera at a frame rate of 30 frames per second. The videos were analyzed by ImageJ, to obtain information on the radius of the GUVs, the center of mass of the spherical unaspirated part of the GUVs, and the center of mass of the particles. A typical trajectory of a particle on a GUV, projected onto the x-y plane, is plotted in Fig.~S2c. The projected trajectory can be converted to a trajectory on the surface of the GUV by:
\begin{equation}
{{z}_{p}}=\sqrt{{{R}_{GUV}^{2}}-{{x}_{p}}^{2}-{{y}_{p}}^{2}}
\end{equation}
where ${R}_{GUV}$ is the radius of the GUV, ${{x}_{p}},{{y}_{p}}$ and ${{z}_{p}}$ are positions of the particle in a Cartesian coordinate system origin from the center of the spherical part of the GUV.  
The displacement of the particle in 3-dimensions is:
\begin{equation}
\Delta r'=\sqrt{{{\left[ \vec{r'}\left( t+\tau  \right)-\vec{r'}\left( t \right) \right]}^{2}}}
\end{equation}
where $\vec{r'}\left( t \right)={{x}_{p}}\left( t \right){{\vec{e}}_{x}}+{{y}_{p}}\left( t \right){{\vec{e}}_{y}}+{{z}_{p}}\left( t \right){{\vec{e}}_{z}}$.
The particle's displacement on the surface of the sphere can be calculated by:
\begin{equation}
\Delta {{r'}_{s}}=2\pi {R}_{GUV}\frac{\beta }{2\pi }
\end{equation}
where $\beta =2\arcsin \left( \frac{\Delta r'}{2{R}_{GUV}} \right)$.
The MSD of a Janus particle is subsequently calculated by this surface displacement and plotted in Fig.~S2d. \\
~\\

textbf{Background flow is not correlated with adhered particle migration}\\
To prove that the Janus particles do not migrate because of weak bulk convection, particles that are not adhered to the GUV were used as tracers for bulk fluid motion. In an experiment where a Janus particle is migrating towards the pipette, we tracked the migrating particle as well as the tracers in the background (Fig.~S3a) to calculate the velocity in the $s$-direction of all particles(Fig.~S3b). No correlation is found between the bulk convection and the migration direction of the particle on the GUV.  \\
~\\

\begin{figure}
\centering
\includegraphics[scale=0.7]{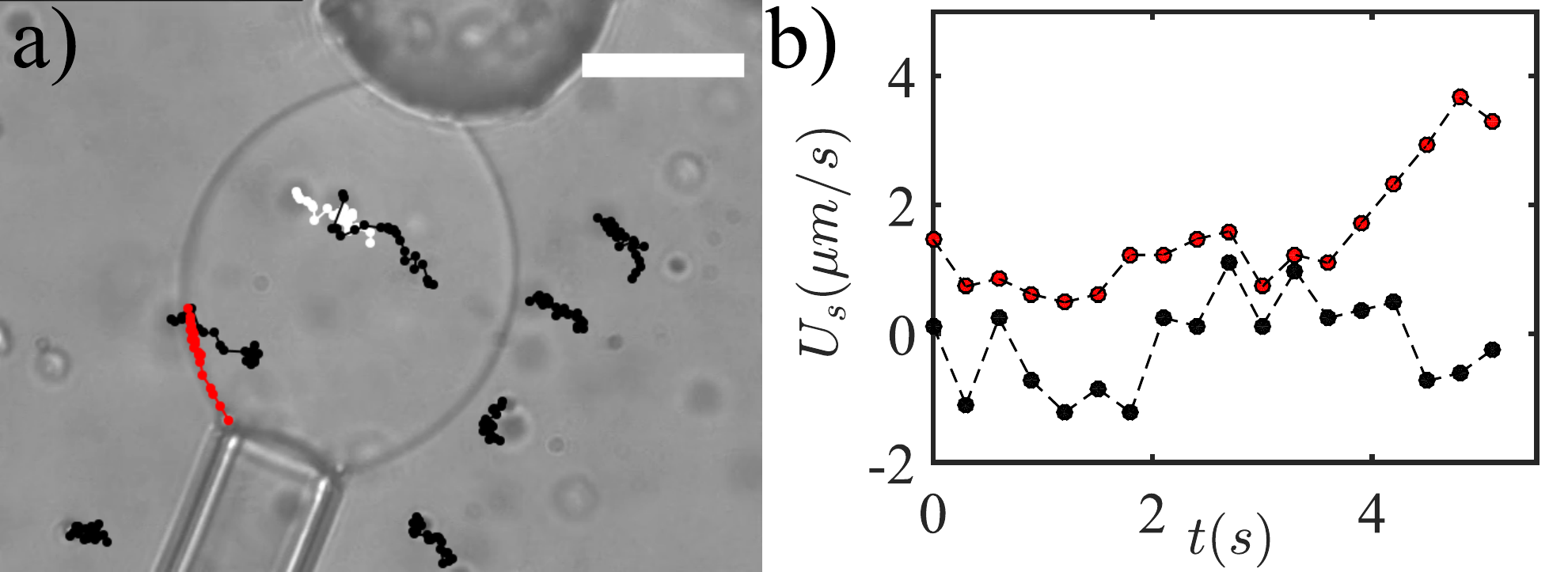}
\begin{addmargin}[.2in]{.2in}
\caption{\footnotesize \textbf{The migration of a Janus particle on an elongated GUV has no correlation to bulk flow} a) Path of a migrating Janus particle (red), and tracers (black and white). b) Velocity in the $s$ direction for the migrating particle (red) and average of the tracers (black). } 

\end{addmargin}

\end{figure}

\textbf{Janus particle's movement on membrane under low tension}\\
Colloids on an elongated vesicle with tension $\sigma=0.05$~$mN/m$ show no significant bias in their motion towards the regions of high curvature over the $60$ second time span of the experiment. This lack of migration can be attributed to a weak capillary force as shown in Fig.~S5. To confirm, we simulated trajectories according to Eq.~2 and 3, assuming ${\sigma}=0.05~mN/m$, ${h}_{qp}=100~nm$,  $D=0.09~{\mu}m^2/s$. The results, plotted in Fig.~S4a, resemble experiments. In this graph, five simulated trajectories are compared to the experimental trajectory in the $s$-direction over a $60$ second time span. Four of the simulated trajectories appear diffusive, with no preferred direction, while one trajectory shows a slight bias towards the high curvature region. The averaged MSD from the simulated trajectories and the experiment are plotted in Fig.~S4b, and have linear regression ${R}^{2}>0.999$. We also simulate trajectories over a 5 minute time span, which is typically far longer in duration compared to our experiments owing to vesicle rupture. In these simulations, the colloids migrate to the high curvature region as shown in Fig.~S4c. This migration is slow, with $Pe \sim 2$ at the region with highest curvature, indicating that the force driving migration is comparable to Brownian forces. \\
~\\
\begin{figure}
\centering
\includegraphics[scale=0.7]{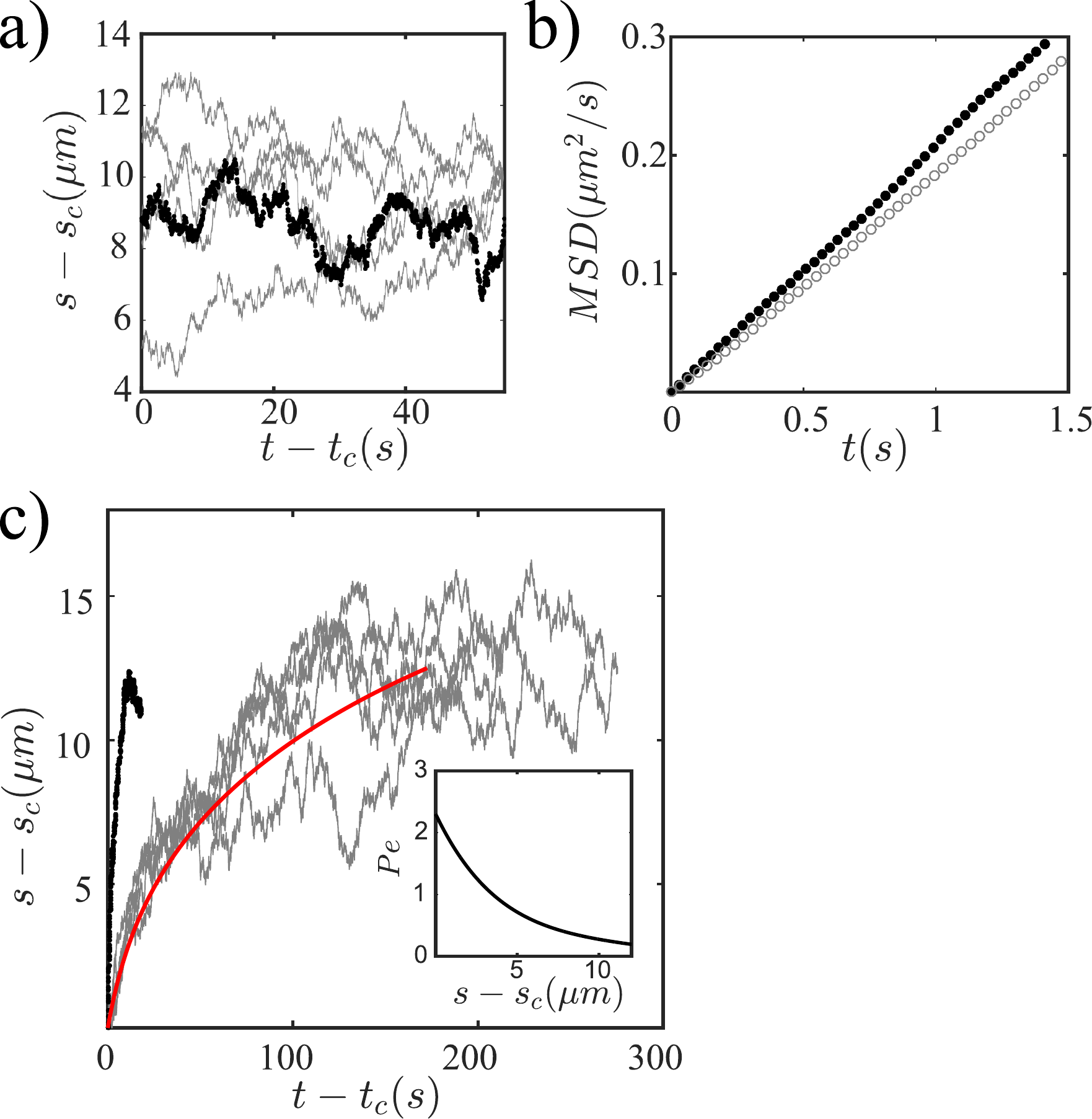}
\begin{addmargin}[.2in]{.2in}
\caption{\footnotesize \textbf{Simulated trajectory of a Janus particles adhered to an elongated GUV at low membrane tension}a) Trajectory in $s$-direction for one experiment (black circles) and 5 simulated realizations (gray lines) for simulation time of $60s$. $\sigma=0.05$~$mN/m$, ${h}_{qp}=100$~$nm$, $D=0.09$~${\mu}m^2/s$. b) MSD in $s$-direction calculated from trajectories in Fig~S13. Black circles: MSD from the experimental trajectory. Gray open circles: Averaged MSD from the simulated trajectories. c) Gray lines: trajectories in $s$-directions for simulations run over 5 minutes. Parameters are the same as those in Fig~S13; Red line: migration under only capillary force with the same conditions. Black circles: high tension experimental data for  $\sigma=0.50$~$mN/m$, ${h}_{qp}=100$~$nm$, $D=0.08$~${\mu}m^2/s$. Inset: $Pe$ number calculated from the capillary migration plotted against migration trajectory in $s$-direction. } 

\end{addmargin}

\end{figure}

\textbf{Capillary force as a function of  tension}\\
We calculated $F$, the capillary force acting on a Janus particle, by differentiation of $\Delta E$ (Eq.~2) for all values of $\sigma$ and GUV shapes in the experiments reported in Fig.~4 in the main text.  In addition, we calculate $F$ for the experiment in which we fixed $\sigma=0.05~mN/m$ and failed to observe capillary migration on an elongated GUV. The results, in Fig.~S5, are normalized by ${k_BT/a}$, the characteristic magnitude of a Brownian force, plotted against particle position on the $s$-coordinate. In this figure, the solid lines indicate the particle positions observed in the experiments. These results indicate that $\sigma =$~$0.05~mN/m$, the capillary force is comparable to Brownian force over the entire elongated GUV. Therefore, no coupling to the curvature of the membrane by the capillary mechanism  is expected at or below this tension.  This is consistent with the observation of diffusive migration at this tension.  Furthermore, it indicates that the threshold tension at which capillary migration becomes unimportant is related to the scale of the particle, the size of the disturbance that it imposes in the membrane, and the shape of the vesicle, which determine the magnitude of $F$ and the characteristic Brownian force. \\
~\\

\begin{figure}
\centering
\includegraphics[scale=0.4]{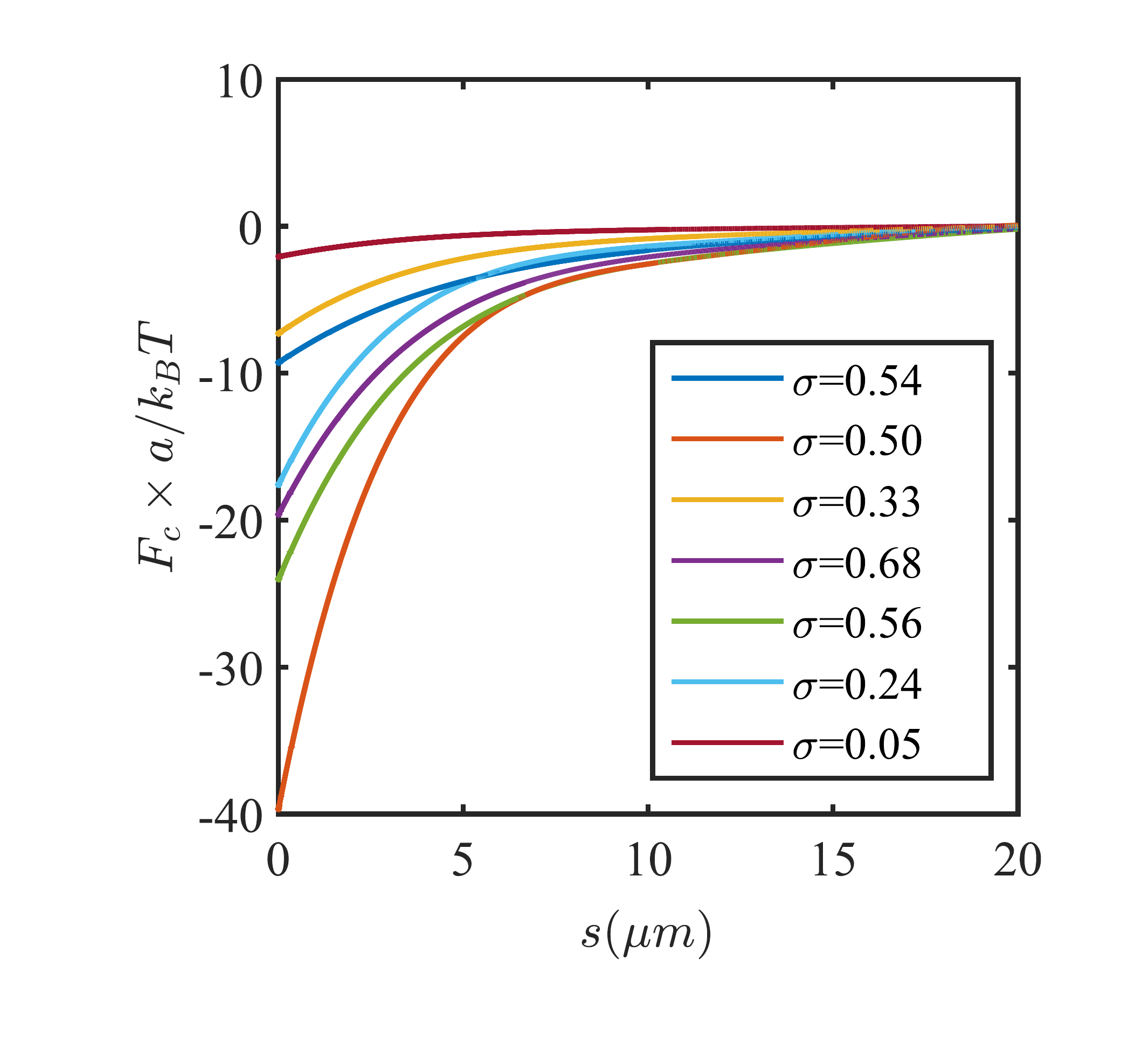}
\begin{addmargin}[.2in]{.2in}
\caption{\footnotesize {Normalized capillary force vs. $s$-position} Force, calculated by differentiation of Eq. 2, for various tension values. GUV shapes and initial particle positions were taken from experiment. The magnitude of the particle-sourced distortion, ${h}_{qp}$ is assumed to be $150~nm$ for all cases. } 

\end{addmargin}

\end{figure}

\textbf{Energy analysis for colloids on tense lipid membranes}

\begin{figure} 
\centering
\includegraphics[scale=0.8]{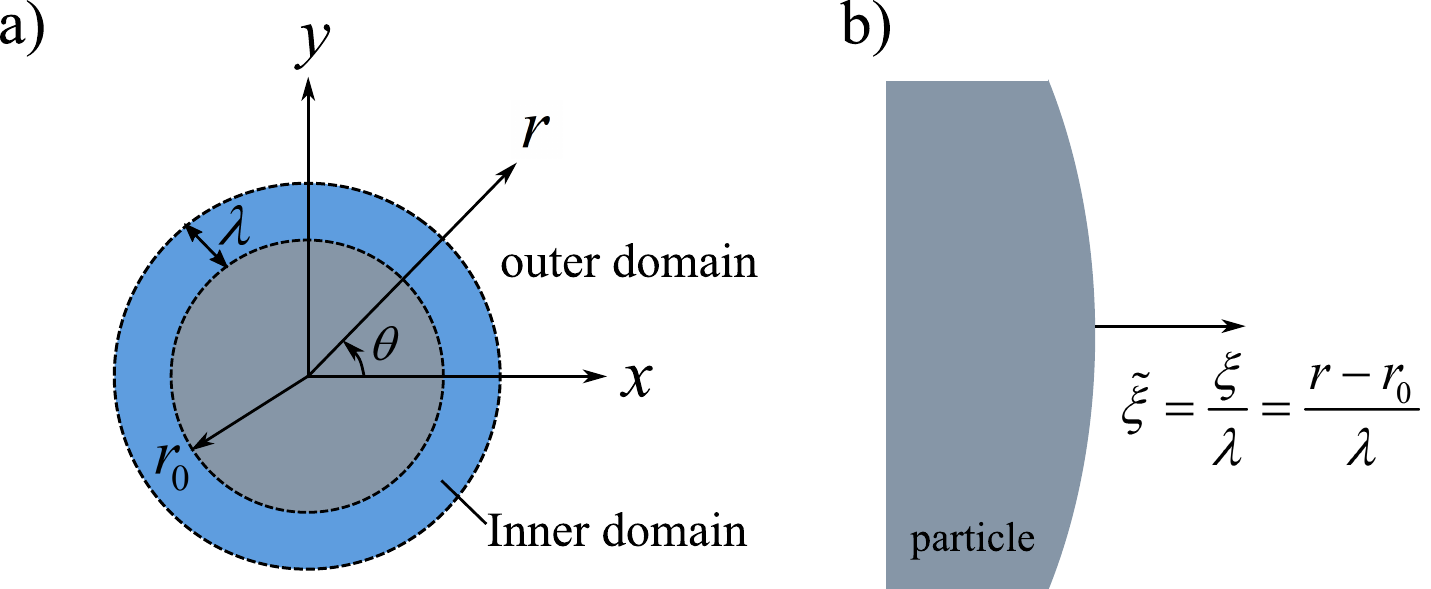}	   
\begin{addmargin}[.2in]{.2in}    
 \caption{\footnotesize \textbf{Schematic of coordinate systems for inner and outer regions }a) Schematic of coordinate system for domain around the adhered particle of radius $r_0$. There is an inner region of thickness $\lambda$ adjacent to the particle, and an outer region where disturbances decay over a distance comparable to $r_0$. b) Stretched inner coordinate measured from particle surface.} 
\end{addmargin}
 \end{figure}

A colloid of radius $a$, adhered to a tense lipid bilayer membrane migrates along membrane curvature gradients to lower the excess membrane area. We define the excess area as the change in membrane area created through particle/membrane interaction. The energy landscape in the context of the Helfrich's model is discussed here.  In this discussion, the ``contact line'', the location where adhered membrane, un-adhered membrane and particle meet, is assumed to be a pinned, undulated contour which can be decomposed into Fourier modes around the location $r=r_0$, where $r_0 \sim a$. In the Helfrich's model, membranes with bending energy $\kappa$ and tension $\sigma$ have an associated natural length scale  $\lambda  = \sqrt {{\kappa  \mathord{\left/ {\vphantom {\kappa  \sigma }} \right. \kern-\nulldelimiterspace} \sigma }}$. Typically, for micron scale particles of contact line radius $r_0$, the ratio $\epsilon={\lambda}/{r_0}\ll 1$. For example, in our experiments, for the range of tension studied, $\epsilon$ varies from 0.02 to 0.04, assuming that $r_0 \sim a$. Below we show that, in this limit, assuming $\left| {\nabla h_0} \right| \ll 1$, the membrane shape is determined by a linearized Young-Laplace equation, with bending energy playing a role only in a small region adjacent to the particle of radial extent  similar to $\lambda$. An asymptotic analysis in small $\epsilon$ shows that the distortion field made by the particle, and the corresponding energy field, reduces, to leading order, to forms reported previously for particles at fluid interfaces. \\
~\\
The shape of the host membrane, absent an adhered colloid, at an arbitrary point $O$ on the membrane is \cite{Liu2015}: 
\begin{eqnarray}
{h_0} = \frac{{{\Delta {c}}}}{4}r^2\cos 2\phi+\frac{H}{2} r^2,
\end{eqnarray}
where $H$ is the mean curvature and $\Delta c$ is deviatoric curvature of the membrane evaluated at $O$, and $(r,\phi)$ is a polar coordinate system with origin $O$. When a particle adheres to this membrane, this representation of the membrane shape is an expansion locally valid in the limit of small $r_0 H$ and small $r_0 \Delta c$ over distances from $O$ comparable to $r_0$.  Since, as we show below, particle-sourced distortions decay within this domain, Eq.~S7 is the far field boundary condition for the membrane shape in the presence of the particle. We are interested in determining the excess area, (the amount of membrane area that is extracted from the area reservoir
represented by the micropipette-aspirated fraction of the vesicle) at mechanical equilibrium in response to particle binding.
Furthermore, we are interested in the free energy change as adhered particles migrate on the membrane for the case where $H$ is constant. In this situation, energetic contributions associated with $H$ are also constant to leading order, and do not drive particle migration along the membrane.  \\
~\\
Here, we find the shape of the membrane after the particle adheres, and subsequently discuss the change in free energy owing to the particle attachment. To solve for the shape of the tense membrane $h$ in the presence of the particle, we begin with Helfrich\textquotesingle s equation:\\
\begin{eqnarray}
\kappa {\nabla ^4}h - \sigma {\nabla ^2}h = \Delta P
\end{eqnarray}
 We recast Eq.~S8 in dimensionless form using $r_0$ as the characteristic length scale:
\begin{eqnarray}
\epsilon ^ 2  {\hat{\nabla} ^4}\hat {h} - {\hat {\nabla} ^2} \hat {h} = \Pi
\end{eqnarray}
where $\hat \nabla  = {r_0}\nabla $, $\hat h  = {h}/{r_0} $ and  $\Pi  = {{\Delta P r_0}}/{\sigma }$. In this form, it is evident that the small parameter multiplies the highest order derivative. To understand the system behavior in the limit of small $\epsilon$, we must divide the region around the particle into two domains. There is an inner region, close to the contact line, where the membrane shape $h^{inner}$ changes over distance comparable to $ \lambda$, and energies associated with membrane bending must balance those associated with tension. There is also an outer region; $h^{outer}$ includes particle sourced disturbances that  decay over distances comparable to $r_0$ in the tense membrane. It is convenient to first discuss the outer region, and subsequently discuss the inner region.\\
~\\
In the outer region, to leading order in $\epsilon$, the membrane shape is governed by:
\begin{eqnarray}
 {\hat \nabla ^2}\hat h^{outer} = -\Pi
\end{eqnarray}
We assume here that the boundary condition at the contact line has no corrections from the inner domain, which we confirm below. For a particle on a tense membrane which has negligible body forces and torques, the leading order mode of the contact line (in outer coordinates) is a  quadrupolar mode with magnitude $h_{qp}$, as shown in the work of Stamou \cite{Stamou2000}.  Thus, $h^{outer}$ obeys the boundary conditions: 
\begin{eqnarray}
 {\hat h^{outer}}({\hat r}=1)={\hat h_{qp}}\cos 2 \phi-\frac{{{r_0 H}}}{2}    
\end{eqnarray}  
and 
\begin{eqnarray}
 {\hat h^{outer}}({\hat r} \to \infty) \to 
{\hat h_0.}
\end{eqnarray}
The solution to Eq.~S10-12 is \cite{Yao2015,Liu2015,Lewandowski2008}: 
\begin{align}
&\hat h^{outer} = \frac{{\Delta {c_0}}{r_0} }{4}{ \hat{r^2}}\cos 2\phi + \frac{1}{{\hat{r^2}}}({\hat{h_{qp}}} - \frac{{r_0\Delta {c_0}}}{4})\cos 2\phi +A_H,
\end{align}
In these expressions, the term ${{{r_0 H}}}/{2}$ in Eq.~S11 appears owing to a shift in the center of mass of the particle, and the term $A_H$ in Eq.~S13 is defined to be $A_H={{r_0 H}}\hat{r^2}/{2}$.  In this latter equation, we have assumed that the distortion at the particle contact line aligns with its quadrupolar rise axis along the rise axis of the host interface.  If these two fields are not aligned, additional terms appear.\cite{Lewandowski2008} To complete the solution to leading order in $\epsilon$, we must determine whether there are corrections to the shape of the membrane owing to the  inner region. There are two cases of boundary conditions admissible for closure of this problem in the inner region, i.e. zero mean curvature at the contact line, or continuity of slopes at the contact line. In both cases, by applying a  Van Dyke matching procedure between the inner and outer regions, we find that there are no corrections to leading order. We present the details below.\\
~\\
We define the inner region variable $\xi $, which measures distance from the contact line so $r = {r_0} + \xi $.  We also define the associated stretched dimensionless variable, $\tilde \xi  = {\xi }/{\lambda }$, so $\hat r = 1 + \epsilon \tilde \xi $.  By recasting $\hat h^{inner}(\tilde \xi, \phi)$, after rescaling and some manipulation, to leading order in $\epsilon$, Eq.~S8 becomes:
\begin{align}
\frac{{{\partial ^4}\hat h^{inner}}}{{\partial {{\tilde \xi }^4}}} - \frac{{{\partial ^2}\hat h^{inner}}}{{\partial {{\tilde \xi }^2}}} = 0
\end{align} with associated general solution:
\begin{align}
\hat h^{inner} = {f_1}(\phi )\exp ( - \tilde \xi ) + {f_2}(\phi )\exp (\tilde \xi ) + {f_3}(\phi )\tilde \xi  + {f_4}(\phi )
\end{align}
where $f_i$, $i=1,2,3,4$ are functions of $\phi$ determined by boundary conditions at the contact line and by matching to the outer solution.  We prepare the outer solution for matching by recasting it in $\tilde \xi$ and retaining only leading order terms:
\begin{align}
\hat h^{outer}(\tilde \xi ,\phi ) = {{\hat h}_{qp}} \cos 2\phi  
\end{align}
This expression matches the inner region if \begin{align}
f_1(\phi ) = {f_2}(\phi ) = {f_3}(\phi ) = 0
\end{align}
and 
\begin{align}
{f_4}(\phi ) = {{\hat h}_{qp}}\cos2\phi 
\end{align}
An admissable solution from the inner region must obey one of the two boundary conditions that emerge from the Euler-Lagrange equations for the Helfrich model. These conditions require that the mean curvature must be zero at the contact line in the inner coordinates, which reduces to 
\begin{align}
{\left( {\frac{{{\partial ^2}\hat h_{}^{inner}}}{{\partial {{\tilde \xi }^2}}}} \right)_{\tilde \xi  = 0}} = 0.
\end{align}
or alternatively, the slope of the membrane must be continuous at the contact line in inner coordinates. Having assumed small slopes, (of order $\epsilon$) the leading order boundary condition for continuity of slopes is: 
\begin{align}
{\left( {\frac{{{\partial }\hat h_{}^{inner}}}{{\partial {{\tilde \xi }}}}} \right)_{\tilde \xi  = 0}} = 0.
\end{align}
Eq.~S18 indeed obeys both Eq.~S19 and S20. This shows that, to leading order,  the membrane shape has no correction from the inner region, and the solution Eq.~S13 is valid to $O(\epsilon)$ everywhere on the membrane. Thus, on tense membranes, bending energy plays no role in the particle-sourced distortion to leading order, and the membrane shape is identical to the solution derived previously for particles on isotropic fluid interfaces with interfacial tension $\sigma$. The associated  energetic consequences are discussed below. \\
~\\
For a tense membrane, absent a colloid, the free energy of the system to leading order in $\epsilon$ is:
\begin{eqnarray}
{E_1} = \sigma \mathop{{\int\!\!\!\!\!\int}\mkern-21mu \bigcirc}\limits_{M} 
 {(1 + \frac{{\nabla {h_0} \cdot \nabla {h_0}}}{2})dS},
\end{eqnarray}
where $dS$ is the area element, $h_0$ is the Monge representation of the host interface,  $\sigma$ is the membrane tension, and the domain $M$ denotes the entire membrane.  This is divided into two regions, where $I$ is the area exterior to the contact line, and $D$ the area in contact with the particle. When the particle interacts with the membrane, adhesion changes the energy over $D$, and particle-sourced distortions alter the energy over $I$.
\begin{eqnarray}
{E_2} = \sigma \mathop{{\int\!\!\!\!\!\int}\mkern-21mu \bigcirc}\limits_I 
 {(1 + \frac{{\nabla h \cdot \nabla h}}{2})dS}  + \frac{\kappa }{2}\mathop{{\int\!\!\!\!\!\int}\mkern-21mu \bigcirc}\limits_I {{{({\nabla ^2}{h})}^2}dS} + {E_0},
\end{eqnarray}
where,
\begin{eqnarray}
{E_0} = \gamma_1(4\pi r_0^2-A_{cap})-\sigma (\pi r_0^2-A_{cap})+{f_{ad}}{A_{cap}} + \frac{\kappa }{{2{r_0^2}}}{A_{cap}},
\end{eqnarray}
In this equation, $\gamma_1$ is the surface energy of the colloid in contact with the surrounding fluid, $A_{cap}$ is the area of membrane adhered to the particle in the form a spherical cap, and $f_{ad}$ ($<0$) is the energy of adhesion per unit area.  The first term is the change in particle surface energy owing to adhesion.  The second term is the change in free energy as area is supplied to form the adhered cap from the area reservoir (the micropipette aspirated fraction). The third term is the energy liberated by particle adhesion, and the fourth term is the energy required to bend the membrane to form $A_{cap}$. The change in free energy owing to the particle adhesion  is:
\begin{align}
&E = {E_2} - {E_1} = {E_0} + \sigma \mathop{{\int\!\!\!\!\!\int}\mkern-21mu \bigcirc}\limits_I 
{(\frac{{\nabla h \cdot \nabla h}}{2} - \frac{{\nabla {h_0} \cdot \nabla {h_0}}}{2})dS} + \frac{\kappa }{2}\mathop{{\int\!\!\!\!\!\int}\mkern-21mu \bigcirc}\limits_I 
 {{{({\nabla ^2}h)}^2}dS}  - \sigma \mathop{{\int\!\!\!\!\!\int}\mkern-21mu \bigcirc}\limits_D 
 {(1+\frac{{\nabla {h_0} \cdot \nabla {h_0}}}{2})dS},\label{enery}
\end{align}
Normalizing Eq.~S24 with $\sigma r_0^2$, an inspection reveals that contributions owing to bending are zero to leading order in $\epsilon$.  In this case, Eq.~S13 and S24 together reduce to the problem of a spherical particle or disk migrating on an isotropic fluid interface owing to capillarity, as discussed in detail in \cite{Yao2015,Liu2015,Mood2016}. The relevant functional form is:
\begin{align}
&E = {E_2} - {E_1} = E_0+ \sigma \pi {r_0^2}h_{qp}^2- \sigma \pi {r_0^2} \frac{3 H^2}{4}- \sigma \pi {r_0^2}\frac{{{h_{qp}}\Delta c(r=0)}}{2}
\end{align}
where the quadratic term in $h_{qp}$ is the self-energy associated with the particle sourced deformation, the quadratic term in $H$ is the $PV$ work to locate the particle center of mass, and the term proportional to ${{h_{qp}}\Delta c(r=0)}$ is the curvature capillary energy driving migration, where $\Delta c(r=0)$ is evaluated at the particle center of mass.  The particle decreases its free energy by moving to sites of largest deviatoric curvature. \\
In our experiments, we consider particles migrating with pinned contact lines on axisymmetric vesicles with fixed tension, constant $H$, and curvature field dependent on meridional arclength $s$. Letting $s_f$ denote the final location along a trajectory, and $s_i$ denote the initial location, we define $\Delta E=E(s_f)-E(s_i)$, and $\Delta C=\Delta c(s_f)-\Delta c(s_i)$, so that the change of free energy as particles migrate on the membrane can be expressed compactly as:
\begin{align}
&\Delta E = - \sigma \pi {r_0^2}\frac{{{h_{qp}}\Delta C}}{2}
\end{align}   
which corresponds to Eq.~2 in the main text. \\
~\\
 
\textbf{Scanning electron microscopy (SEM)}\\
A droplet of the particle suspension was dried on a cover slip. A thin layer (<10 $nm$) of a palladium and gold alloy was deposited on the sample by sputtering (Sputter Coater 108, Cressington Scientific, UK). The sputtered sample was imaged by a scanning electron microscope (Quanta 600 ESEM, FEI Corporations, Hillsboro, OR). \\
~\\

\textbf{Zeta potential measurement}\\
Zeta potential measurements for all particles used in this study were obtained using DelsaNano C analyzer, (Beckman Coulter, Inc., Brea, CA). Particles suspensions were diluted roughly 100 fold from a stock concentration of 1 $\%$ solid to be sufficiently dilute for detection of particle displacement under the applied field within the analyzer. The zeta potential is $-38.6 \pm 1.2$~$mV$ for the Janus colloids,  and $-56.2 \pm 3.0$~$mV$ for the PS colloids. \\
~\\

\textsc{\textbf{Atomic force microscopy (AFM)}}\\
AFM was performed on dried particles to investigate the surface roughness. A drop of the suspension of particles was dried on a piece of flat PDMS, and imaged in the tapping mode (Dimension Icon AFM, Bruker, Billerica, MA). For AFM measurements in a water environment, dry particles were deposited on a thin SU8 layer around 500 $nm$, and the SU8 was cured by UV light. The particles thus fixed in the SU8 film were rehydrated by DI-water, and imaged in contact mode in water (Asylum Reseach, Santa Barbara, CA). AFM images were post-processed using a open source software Gwyddion. A quadratic function was fitted and subtracted from the particle surface, and a root mean square (RMS) of the height of the flattened particle surface was measured and tabulated in table S1. \\

\begin{table}
\centering
\begin{tabular}{| c | c | c |}
    \hline
   \hspace{.1in}Environment\hspace{.1in} & \hspace{.1in}Region on the particle \hspace{.1in} &\hspace{.1in}RMS roughness (nm)\hspace{.1in}  \\   \hline
Air&Smooth PAA side&1.89$\pm$0.8 \\ \hline
Air&Rough PAA/PS side&13.3$\pm$3.1 \\ \hline
Air&Between two sides&21.9\\ \hline
Water&Rough PAA/PS side&5.67$\pm$0.6\\ \hline
    \end{tabular}
\caption{\textbf{RMS roughness on the surface of Janus particle as measured by AFM}}
\end{table}

\begin{figure}
\centering
\includegraphics[scale=0.8]{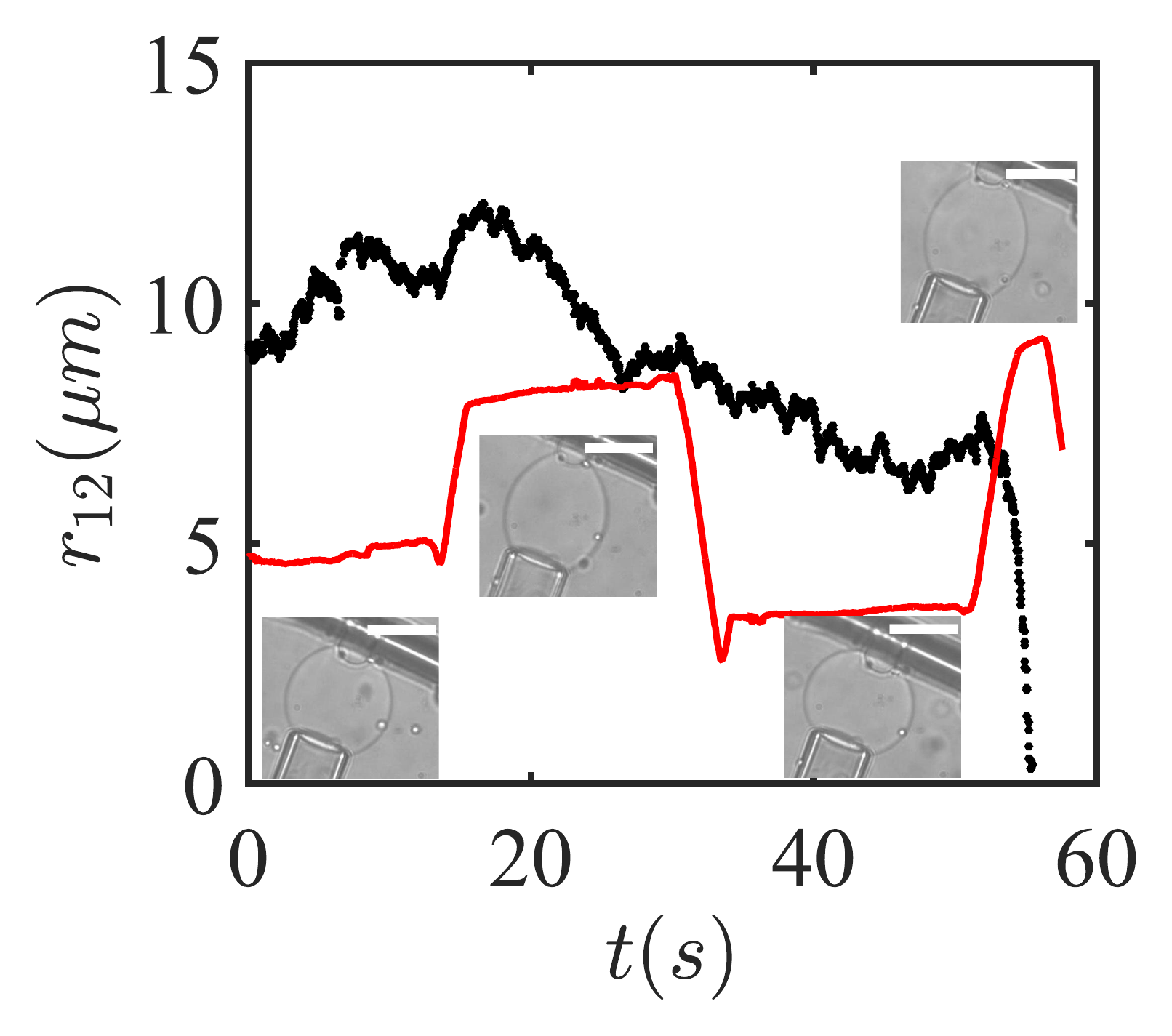}
\begin{addmargin}[.2in]{.2in}
\caption{\footnotesize {Stop-and-go behavior of a Janus particle} Reference position: position were the Janus particle is in contact with the glass bead. Red line: Distance between the pipette with the glass bead and the reference position. As shown in the inset, the GUV is elongated when the pipette is moved away from the reference position. Scale bar: 15 $\mu m$ Black dots: projected distance from contact of the Janus particle. When the GUV was elongated, the particle moved towards the bead. When the curvature gradients were removed, the particle lost the bias in migration direction. This process can be turned ``on'' and ``off'' by changing the GUV shape.  } 

\end{addmargin}

\end{figure}
\pagebreak

\bibliographystyle{unsrt}
\bibliography{curvature_driven_migration_of_colloids_on_lipid_bilayers}

\begin{thebibliography}{10}

\bibitem{Sorre2011}
A~Callan-Jones, B~Sorre, and P~Bassereau.
\newblock Curvature-driven sorting in biomembranes.
\newblock {\em Cold Spring Harb Perspect Biol}, 3(2), 2011.

\bibitem{Zhu2011}
T~Baumgart, BR~Capraro, C~Zhu, and SL~Das.
\newblock Thermodynamics and mechanics of membrane curvature generation and
  sensing by proteins and lipids.
\newblock {\em Annu Rev Phys Chem}, 62(1):483, 2011.

\bibitem{Tian2009}
A~Tian and T~Baumgart.
\newblock Sorting of lipids and proteins in membrane curvature gradients.
\newblock {\em Biophys J}, 96(7):2676--2688, 2009.

\bibitem{Capraro2010}
BR~Capraro, Y~Yoon, W~Cho, and T~Baumgart.
\newblock Curvature sensing by the epsin n-terminal homology domain measured on
  cylindrical lipid membrane tethers.
\newblock {\em J Am Chem Soc}, 132(4):1200--201, 2010.

\bibitem{Mitragotri2015}
AC~Anselmo, S~Kumar, V~Gupta, A~Pearce, A~Ragusa, V~Muzykantov, and
  S~Mitragotri.
\newblock Light-controlled topological charge in a nematic liquid crystal.
\newblock {\em Biomaterials}, 68:1--8, 2015.

\bibitem{Petros2010}
RA~Petros and JM~DeSimone.
\newblock Strategies in the design of nanoparticles for therapeutic
  applications.
\newblock {\em Nat Rev Drug Discov}, 9(8):615--627, 2010.

\bibitem{Safinya1999}
I~Koltover, JO~Raedler, and CR~Safinya.
\newblock Membrane mediated attraction and ordered aggregation of colloidal
  particles bound to giant phospholipid vesicles.
\newblock {\em Phys Rev Lett}, 82(9):1991, 1999.

\bibitem{Beales2012}
S~Zhang, A~Nelson, and PA~Beales.
\newblock Freezing or wrapping: the role of particle size in the mechanism of
  nanoparticle–biomembrane interaction.
\newblock {\em Langmuir}, 28(35):12831--12837, 2012.

\bibitem{Fery2003}
A~Fery, S~Moya, PH~Puech, F~Brochard-Wyart, and H~Mohwald.
\newblock Interaction of polyelectrolyte coated beads with phospholipid
  vesicles.
\newblock {\em C R Phys}, 4(2):259--264, 2003.

\bibitem{Weikl2014}
AH~Bahrami, M~Raatz, J~Agudo-Canalejo, R~Michel, EM~Curtis, CK~Hall,
  M~Gradzielski, R~Lipowsky, and TR~Weikl.
\newblock Wrapping of nanoparticles by membranes.
\newblock {\em Adv Colloid Interface Sci}, 208:214--224, 2014.

\bibitem{Tu2014}
F~Tu and D~Lee.
\newblock Shape-changing and amphiphilicity-reversing janus particles with
  ph-responsive surfactant properties.
\newblock {\em J Am Chem Soc}, 136(28):9999--100006, 2014.

\bibitem{Saric2012}
A~\u{S}ari\'{c} and A~Cacciuto.
\newblock Fluid membranes can drive linear aggregation of adsorbed spherical
  nanoparticles.
\newblock {\em Phys Rev Lett}, 108(11):118101, 2012.

\bibitem{Cavallaro2011}
M~Cavallaro, L~Botto, EP~Lewandowski, M~Wang, and KJ~Stebe.
\newblock Curvature-driven capillary migration and assembly of rod-like
  particles.
\newblock {\em Proc Natl Acad Sci}, 108(52):20923--20928, 2011.

\bibitem{Liu2015}
N~Sharifi-Mood, IB~Liu, and KJ~Stebe.
\newblock Curvature capillary migration of microspheres.
\newblock {\em Softmatter}, 11(34):6768--6779, 2015.

\bibitem{Mood2016}
N~Sharifi-Mood, IB~Liu, and KJ~Stebe.
\newblock Reply to the comments on "curvature capillary migration of
  microspheres" by p. galatola and a. w{\"u}rger. soft matter.
\newblock {\em Softmatter}, 12(2):333--336, 2016.

\bibitem{Yu2009}
Y~Yu and S~Granick.
\newblock Pearling of lipid vesicles induced by nanoparticles.
\newblock {\em J Am Chem Soc}, 131(40):14158--14159, 2009.

\bibitem{McMahon2005}
HT~McMahon and JL~Gallop.
\newblock Membrane curvature and mechanisms of dynamic cell membrane
  remodelling.
\newblock {\em Nature}, 438(7068):590--596, 2005.

\bibitem{Simunovic2015}
M~Simunovic, GA~Voth, A~Callan-Jones, and P~Bassereau.
\newblock ematic liquid crystals.
\newblock {\em Trends Cell Bio}, 25(12):780--792, 2015.

\bibitem{Stamou2000}
D~Stamou, C~Duschl, and D~Johannsmann.
\newblock Long-range attraction between colloidal spheres at the air-water
  interface: The consequence of an irregular meniscus.
\newblock {\em Phys Rev E}, 62:5263--5272, 2000.

\bibitem{Loudet2005}
JC~Loudet, AM~Alsayed, J~Zhang, and AG~Yodh.
\newblock Capillary interactions between anisotropic colloidal particles.
\newblock {\em Phys Rev Lett}, 94:018301.

\bibitem{Lewandowski2008}
EP~Lewandowski, JA~Bernate, PC~Searson, and KJ~Stebe.
\newblock Rotation and alignment of anisotropic particles on nonplanar
  interfaces.
\newblock {\em Langmuir}, 24:9302--9307, 208.

\bibitem{Ershov2013}
D~Ershov, J~Sprakel, J~Appel, MAC Stuart, and J~van~der Gucht.
\newblock Capillarity-induced ordering of spherical colloids on an interface
  with anisotropic curvature.
\newblock {\em Proc Natl Acad Sci}, 110(23):9220--9224, 2013.

\bibitem{Mathivet1996}
L~Mathivet, S~Cribier, and PF~Devaux.
\newblock Shape change and physical properties of giant phospholipid vesicles
  prepared in the presence of an ac electric field.
\newblock {\em Biophys J}, 70:1112--21, 1996.

\bibitem{Doyle2005}
T~Savin and PS~Doyle.
\newblock Static and dynamic errors in particle tracking microrheology.
\newblock {\em Biophys J}, 88(1):623--638, 2005.

\bibitem{Domanov2011}
YA~et~al. Domanov.
\newblock Mobility in geometrically confined membranes.
\newblock {\em Proc Natl Acad Sci}, 108(31):12605--12610, 2011.

\bibitem{Stone2015}
HA~Stone and H~Masoud.
\newblock Mobility of membrane-trapped particles.
\newblock {\em J Fluid Mech}, 781:494--505, 2015.

\bibitem{Yao2015}
L~Yao, N~Sharifi-Mood, IB~Liu, and KJ~Stebe.
\newblock Capillary migration of microdisks on curved interfaces.
\newblock {\em J Colloid Inrerface Sci}, 449:436--442, 2015.

\end{thebibliography}

\end{document}